\documentclass{article}

    \PassOptionsToPackage{numbers, compress}{natbib}
\usepackage[preprint]{neurips_2022}

\usepackage[utf8]{inputenc} % allow utf-8 input
\usepackage[T1]{fontenc}    % use 8-bit T1 fonts
\usepackage{xr}
\usepackage{xr-hyper}
\usepackage{hyperref}       % hyperlinks
\usepackage{url}            % simple URL typesetting
\usepackage{booktabs}       % professional-quality tables
\usepackage{amsfonts}       % blackboard math symbols
\usepackage{microtype}      % microtypography
\usepackage{xcolor}         % colors
\usepackage{amsmath}
\usepackage{authblk} 
\usepackage[version=4]{mhchem}
\usepackage{multirow}
\usepackage{subfigure}
\usepackage{graphicx}
\usepackage{bm}
\usepackage{siunitx}
\usepackage[normalem]{ulem}

\usepackage{multirow} % for cmd 'multirow', 'multicolumn'
\usepackage{array}
\newcommand{\PreserveBackslash}[1]{\let\temp=\\#1\let\\=\temp}
\newcolumntype{C}[1]{>{\PreserveBackslash\centering}p{#1}}
\newcolumntype{R}[1]{>{\PreserveBackslash\raggedleft}p{#1}}
\newcolumntype{L}[1]{>{\PreserveBackslash\raggedright}p{#1}}

\usepackage{soul}
\usepackage{color}

\newcommand{\wh}[1]{{\color{black}#1}}

\newcommand{\rr}{h}
\newcommand{\hh}{g}

\makeatletter
\newcommand*{\addFileDependency}[1]{% argument=file name and extension
  \typeout{(#1)}
  \@addtofilelist{#1}
  \IfFileExists{#1}{}{\typeout{No file #1.}}
}
\makeatother

\title{DPA-2: a large atomic model as a multi-task learner}

\author[1,2,3]{\textbf{Duo Zhang\thanks{These authors contributed equally to this work.}~~}}
\author[1,2]{\textbf{Xinzijian Liu}\protect\footnotemark[1]~~}
\author[4,5]{\textbf{Xiangyu Zhang}}
\author[2,6]{\textbf{Chengqian Zhang}}
\author[1,2]{\textbf{Chun Cai}}
\author[1,2]{\textbf{Hangrui Bi}}
\author[4,5]{\textbf{Yiming Du}}
\author[7,8]{\textbf{Xuejian Qin}}
\author[1]{\textbf{Anyang Peng}}
\author[2,9]{\textbf{Jiameng Huang}}
\author[10]{\textbf{Bowen Li}}
\author[7,8]{\textbf{Yifan Shan}}
\author[11]{\textbf{Jinzhe Zeng}}
\author[2]{\textbf{Yuzhi Zhang}}
\author[2]{\textbf{Siyuan Liu}}
\author[12]{\textbf{Yifan Li}}
\author[2,13]{\textbf{Junhan Chang}}
\author[2]{\textbf{Xinyan Wang}}
\author[2,14]{\textbf{Shuo Zhou}}
\author[15]{\textbf{Jianchuan Liu}}
\author[16,17]{\textbf{Xiaoshan Luo}}
\author[17,18]{\textbf{Zhenyu Wang}}
\author[1]{\textbf{Wanrun Jiang}}
\author[19]{\textbf{Jing Wu}}
\author[19]{\textbf{Yudi Yang}}
\author[19]{\textbf{Jiyuan Yang}}
\author[20]{\textbf{Manyi Yang}}
\author[21]{\textbf{Fu-Qiang Gong}}
\author[2]{\textbf{Linshuang Zhang}}
\author[2]{\textbf{Mengchao Shi}}
\author[1]{\textbf{Fu-Zhi Dai}}
\author[11]{\textbf{Darrin M. York}}
\author[19,22]{\textbf{Shi Liu}}
\author[10,23,24]{\textbf{Tong Zhu}}
\author[7,8]{\textbf{Zhicheng Zhong}}
\author[17]{\textbf{Jian Lv}}
\author[21,25,26]{\textbf{Jun Cheng}}
\author[4]{\textbf{Weile Jia}}
\author[1,6]{\textbf{Mohan Chen}}
\author[2]{\textbf{Guolin Ke}}
\author[1,27,28]{\textbf{Weinan E}}
\author[1,2, \thanks{\href{mailto:linfeng.zhang.zlf@gmail.com}{linfeng.zhang.zlf@gmail.com}}]{\textbf{Linfeng Zhang}}
\author[6,29, \thanks{\href{mailto:wang_han@iapcm.ac.cn}{wang\_han@iapcm.ac.cn}}]{\textbf{Han Wang}}
\affil[1]{AI for Science Institute, Beijing 100080, P. R.~China}
\affil[2]{DP Technology, Beijing 100080, P. R.~China}
\affil[3]{Academy for Advanced Interdisciplinary Studies, Peking University, Beijing 100871, P. R.~China}
\affil[4]{State Key Lab of Processors, Institute of Computing Technology, Chinese Academy of Sciences, Beijing 100871, P.R.~China}
\affil[5]{University of Chinese Academy of Sciences, Beijing 100871, P.R.~China}
\affil[6]{HEDPS, CAPT, College of Engineering, Peking University, Beijing 100871, P.R.~China}
\affil[7]{Ningbo Institute of Materials Technology and Engineering, Chinese Academy of Sciences, Ningbo 315201, P.R.~China}
\affil[8]{CAS Key Laboratory of Magnetic Materials and Devices and Zhejiang Province Key Laboratory of Magnetic Materials and Application Technology, Chinese Academy of Sciences, Ningbo 315201, P.R.~China}
\affil[9]{School of Electronics Engineering and Computer Science, Peking University,
 Beijing 100871, P.R.~China}
\affil[10]{Shanghai Engineering Research Center of Molecular Therapeutics \& New Drug Development, School of Chemistry and Molecular Engineering, East China Normal University, Shanghai 200062, P.R.~China}
\affil[11]{Laboratory for Biomolecular Simulation Research, Institute for Quantitative Biomedicine and Department of Chemistry and Chemical Biology, Rutgers University, Piscataway, New Jersey 08854, USA}
\affil[12]{Department of Chemistry, Princeton University, Princeton, New Jersey 08540, USA}
\affil[13]{College of Chemistry and Molecular Engineering, Peking University, Beijing 100871, P.R.~China}
\affil[14]{Yuanpei College, Peking University, Beijing 100871, P.R.~China}
\affil[15]{School of Electrical Engineering and Electronic Information, Xihua University, Chengdu, 610039, P.R.~China}
\affil[16]{State Key Laboratory of Superhard Materials, College of Physics, Jilin University, Changchun 130012, P.R.~China}
\affil[17]{Key Laboratory of Material Simulation Methods \& Software of Ministry of Education, College of Physics, Jilin University, Changchun, 130012, P.R.~China}
\affil[18]{International Center of Future Science, Jilin University, Changchun, 130012, P.R.~China}
\affil[19]{Key Laboratory for Quantum Materials of Zhejiang Province, Department of Physics, School of Science, Westlake University, Hangzhou, Zhejiang 310030, P.R.~China}
\affil[20]{Atomistic Simulations, Italian Institute of Technology, 16156 Genova, Italy}
\affil[21]{State Key Laboratory of Physical Chemistry of Solid Surface, iChEM, College of Chemistry and Chemical Engineering, Xiamen University, Xiamen, 361005, P.R.~China}
\affil[22]{Institute of Natural Sciences, Westlake Institute for Advanced Study, Hangzhou, Zhejiang 310030, P.R.~China}
\affil[23]{NYU-ECNU Center for Computational Chemistry at NYU Shanghai, Shanghai 200062, P.R.~China}
\affil[24]{Institute for Advanced algorithms research, Shanghai, 201306, P.R.~China}
\affil[25]{Laboratory of AI for Electrochemistry (AI4EC), IKKEM, Xiamen, 361005, P.R.~China}
\affil[26]{Institute of Artificial Intelligence, Xiamen University, Xiamen, 361005, P.R.~China}
\affil[27]{Center for Machine Learning Research, Peking University, Beijing 100871, P.R.~China}
\affil[28]{School of Mathematical Sciences, Peking University, Beijing, 100871, P.R.~China}
\affil[29]{Laboratory of Computational Physics, Institute of Applied Physics and Computational Mathematics, Fenghao East Road 2, Beijing 100094, P.R.~China}

\begin{document}
\maketitle

% \newpage
\begin{abstract}

%The development of machine learning-based potential energy surface (PES) models has significantly advanced the field of molecular simulations. As data and models accumulate, there have been preliminary attempts to explore pre-training schemes. However, to truly achieve this new paradigm, which involves upstream pre-training and downstream task finetuning with minimal data, there are many challenges that must be addressed. In this work, we further systematize the pre-train-then-finetune paradigm for machine learning potential energy functions and propose \textbf{transferability} as the most important evaluation criterion. We test the overall capabilities of existing model frameworks under this paradigm on numerous application datasets and discover important characteristics that may be easily overlooked when training on individual datasets but are critical in this context. Furthermore, we introduce a new model architecture, DPA-2, which is better suited for this paradigm and demonstrates improved accuracy and, more importantly, transferability. To achieve an even more unified model, we simultaneously train a single DPA-2 on eight distinct downstream systems using a multi-task training approach, resulting in a transferable model that aligns with the development of this paradigm. Our experiments confirm the excellent transferability of this unified model.

The rapid advancements in artificial intelligence (AI) are catalyzing transformative changes in atomic modeling, simulation, and design. AI-driven potential energy models have demonstrated the capability to conduct large-scale, long-duration simulations with the accuracy of {\it ab initio} electronic structure methods. However, the model generation process remains a bottleneck for large-scale applications.
We propose a shift towards a model-centric ecosystem, wherein a large atomic model (LAM), pre-trained across multiple disciplines, can be efficiently fine-tuned and distilled for various downstream tasks, thereby establishing a new framework for molecular modeling.
In this study, we introduce the DPA-2 architecture as a prototype for LAMs. Pre-trained on a diverse array of chemical and materials systems using a multi-task approach, DPA-2 demonstrates superior generalization capabilities across multiple downstream tasks compared to the traditional single-task pre-training and fine-tuning methodologies. Our approach sets the stage for the development and broad application of LAMs in molecular and materials simulation research.

% This article presents the development of the DP model system from the perspective of Machine Learning Operations (MLOps), emphasizing the importance of incorporating MLOps principles to streamline the model lifecycle, encompassing development, deployment, and maintenance stages. 
% By sharing our experiences and providing an overview of the current state of DP model development, we seek to promote the adoption of MLOps principles within the molecular and materials simulation community, thereby fostering increased collaboration and innovation in this rapidly advancing field. 
% Specifically, we identify several key factors that may be overlooked by the community, which could be critical for addressing challenges related to the escalating complexity of machine learning models and simulations. 
% This ultimately advocates for a more systematic approach to model development, validation, and deployment.

\end{abstract}

\section{Introduction}

%\WH{Importance of MLPs}
An accurate interatomic potential energy surface (PES) is crucial for molecular modeling and simulations. Quantum mechanical (QM) methods, such as density functional theory (DFT)~\cite{hohenberg1964inhomogeneous,kohn1965self}, provide satisfactory accuracy in most applications. 
However, their computational complexity typically scales as the cubic order of the system size, thus limiting large-scale simulations. 
In contrast, empirical force fields (EFF) are way more efficient, but their accuracy is often deemed insufficient for various applications.
Machine learning potentials (MLPs) have emerged as a powerful approach to modeling complex materials and molecules, bridging the gap between the high accuracy of QM methods and the computational efficiency of EFFs. 
This has enabled the study of large-scale molecular systems with QM-level accuracy across diverse applications, including drug discovery~\cite{badaoui2022combined,zeng2023qdpi}, materials design~\cite{bartok2018machine,deringer2020general,wen2022deep}, and catalysis~\cite{ma2020machine,yang2022ammonia}, etc.

%\WH{limitations of the traditional MLPs}
In most MLP applications, the training data is generated from scratch either through brute force \emph{ab initio} molecular dynamics simulations~\cite{car1985unified} or by using a concurrent learning (or active learning) scheme capable of automatically generating the most critical data for building uniformly accurate models~\cite{smith2018less,uteva2018active,zhang2019active,zhang2020dp}.
% Alternatively, the MLP can be trained using an active learning procedure that constructs an optimal subset from a large proposed dataset~\cite{smith2018less,uteva2018active}.
In any case, DFT-calculated energies and forces are required for each configuration in the training dataset, resulting in a substantial amount of efforts spent on constructing DFT-labeled datasets.
For instance, in the AlMgCu general-purpose ternary alloy MLP~\cite{jiang2021accurate}, more than 10 million CPU hours were spent on labeling the 141K training data points.
Furthermore, MLPs often struggle to generalize to applications not covered by the training data~\cite{bartok2018machine}, such as when additional elements are included in materials design or when crystal structures in a broader range of thermodynamic conditions need to be explored.

To further extend the application range of MLPs, efforts have been made to develop ``universal'' or ``fundamental'' models~\cite{takamoto2022towards,chen2022universal,choudhary2023unified,deng2023chgnet,batatia2023foundation,merchant2023scaling}, referred to as large atomic models (LAMs), based on extensive density functional theory (DFT)-labeled datasets. However, the technical approach still requires further exploration, and a LAM-centric ecosystem remains to be established. The primary factors influencing this exploration process are the methods employed for model training and their subsequent application in various tasks.

During the model training stage, a single-task-based training strategy, i.e., training using consistently labeled data, remains dominant. 
Models generated in this way are typically expected to be directly applicable to downstream tasks in which the explored configurations are effectively covered by the training data.
Some examples include models such as M3GNet~\cite{chen2022universal}, CHGNet~\cite{deng2023chgnet} and MACE-MP-0~\cite{batatia2023foundation}, which are all trained on snapshots from DFT relaxations of the Material Project \cite{jain2013materials} structures, with M3GNet utilizing 88K configurations across 89 chemical species and both CHGNet and MACE-MP-0 being trained on 1.58M inorganic crystal frames from the concurrently introduced MPtrj dataset~\cite{deng2023chgnet}; 
GNoME~\cite{merchant2023scaling}, trained on a dataset of inorganic crystals also starting from MP, but nearly two orders of magnitude larger than MPtrj;
PreFerred Potential (PFP), trained on approximately 9M frames of 45 elements~\cite{takamoto2022towards}; 
and ALIGNN, trained on 307K data frames of 89 elements~\cite{choudhary2023unified}. 

Several limitations exist in the single-task training strategy: 
(1) Simultaneously training multiple datasets from different application fields is not feasible due to the variations in labeling with different DFT settings.
For instance, the MPtrj dataset, labeled by DFT calculations using PBE/PBE+U~\cite{perdew1996generalized} exchange-correlation functional and plane-wave basis, cannot be concurrently trained with the ANI-1x dataset, labeled by DFT calculations using the $\omega$B97x hybrid functional~\cite{chai2008systematic} and an atomic basis set, thus little possibility is left to improve the model's generalizability on molecular applications. 
% datasets labeled with different DFT settings, such as exchange-correlation functionals, basis sets, and k-space sampling, cannot be used simultaneously to enhance the model's capability; 
(2) The requirements of downstream tasks might be difficult to satisfy. For instance, a task may require DFT accuracy at the meta-general gradient approximation (meta-GGA) level. 
A model trained with GGA-level DFT data would not be easily adapted to fulfill this requirement.
% Overall, this approach has served well to make progress on relatively narrow domains, but it can hardly be expected to result in a truly new and universal starting point for atomic modeling tasks.

% On the other hand, as discovered in the field of large language models (LLM), the prevalence of single-task training on single-domain datasets may be a major contributor to the lack of generalization observed in current systems. 
% This is evident in the case of GPT-2~\cite{radford2019language}, a model system preceding the era of LLMs. 
Multi-task pre-training, combined with various strategies for downstream tasks such as fine-tuning and distillation, has emerged as a promising alternative for the development of LAMs~\cite{cui2023gpip,feng2023may,jacobson2023leveraging,wang2023denoise}. 
By employing the multi-task training strategy~\cite{kokkinos2017ubernet,devlin2018bert}, it becomes possible to jointly pre-train models using multiple datasets labeled with different DFT settings~\cite{jacobson2023leveraging,shoghi2023molecules}. 
During fine-tuning for downstream tasks, the model's backbone, which encodes the representation of configurational and chemical spaces, is preserved and connected to one or multiple task heads~\cite{smith2019approaching,kolluru2022transfer}. 
As a result, the labeling methods for pre-training and fine-tuning datasets do not need to be identical. Furthermore, the downstream tasks can involve property predictions rather than PES modeling~\cite{shoghi2023molecules}. This scheme offers significant flexibility in downstream tasks and may lead to a much better generalization ability of a LAM.

%\WH{Requirements in terms of model construction, training strategy and validation for pre-training-finetuning.}
Before proceeding further, let us list the requirements of a LAM that we consider to be fundamental:
(1) highly generalizable,
(2) extensive and respect the translational, rotational, and permutational symmetries,
(3) conservative, and
(4) continuous up to second-order derivatives.
A model with high generalizability implies that when trained with the same amount of data, the model can achieve high accuracy~\cite{goodfellow2016deep}.
The generalizability is critical in pre-training LAMs, considering that the DFT-labeled data are expensive and sparse in the configurational and chemical spaces.
% In other words, less data is needed to achieve the same level of accuracy.
% High generalizability would help improve the quality of the LAM model, considering that the DFT-labeled data are expensive and limited.
% The LAM-PES model is extensive indicates that the energy scales linearly with the size of the system.
By conservative, we mean that the forces (and virial tensor, for periodic systems) are calculated by the derivatives of the model-predicted total energy of the system concerning atom coordinates (and cell tensor, respectively).
The conservativeness and smoothness of the model are critical for energy conservation in MD simulations and are thus a compulsory requirement for calculating dynamic properties such as diffusion coefficient, viscosity, and thermal conductivity~\cite{tuckerman2010statistical}.
The requirements (1)--(4) are physical restraints imposed on a PES, thus they are necessary (but in general not sufficient) conditions for the generalizability of the LAMs.
%\WH{Suitable for massively parallel MD simulations.}

%\WH{The contribution of this work} 
%1. Proposed DPA-2 model.
%2. LAM-DPA-2 model for molecular simulations. 
%3. Pre-training, finetuning, distillation workflow for applications. 

\wh{In this context, the primary contribution of this work is the development of DPA-2, a multi-task pre-trained model that meets all the mentioned requirements and furnishes a representation suitable for a diverse array of multi-disciplinary applications, including alloys, semiconductors, battery materials, drug molecules, and more, while exhibiting a high degree of generalization for downstream tasks.
The revelation of a remarkable correspondence between the learned representations by DPA-2 and existing chemical knowledge underscores the potential of the proposed model architecture and the multi-task training scheme. 
Furthermore, we emphasize the importance of an open and application-oriented model evaluation system for the molecular simulation community in the era of large atomic models.
}

% In this context, this work aims to propose DPA-2, a novel architecture for a LAM, and to develop a comprehensive pipeline for model pre-training, fine-tuning, distillation, and application, associated with automatic workflows.
% Although a hybrid multi-task pre-training approach using both labeled and unlabeled data is technically feasible, we focus on supervised learning for pre-training in this work, and leave the investigation of hybrid multi-task pre-training in future studies.
% We demonstrate that with a significantly improved model architecture, the DPA-2 is advantagous over state-of-the-art models like the Gemnet-OC~\cite{gasteiger2022graph}, Equiformer-V2~\cite{liao2023equiformerv2}, NequIP~\cite{batzner20223} and Allegro~\cite{musaelian2022learning} in single-task training and genernalization benchmarks.
% More importantly, DPA-2 can be pre-trained using various datasets covering a broad range of application domains, such as alloys, semi-conductors, battery materials, and drug molecules, etc., with a high generalization ability for downstream tasks. 
% Furthermore, through several examples, we highlight and emphasize that an open and application-oriented model evaluation system is crucial to the molecular simulation community in the era of large atomic models.

\begin{figure*}
    \includegraphics[width=1.0\textwidth]{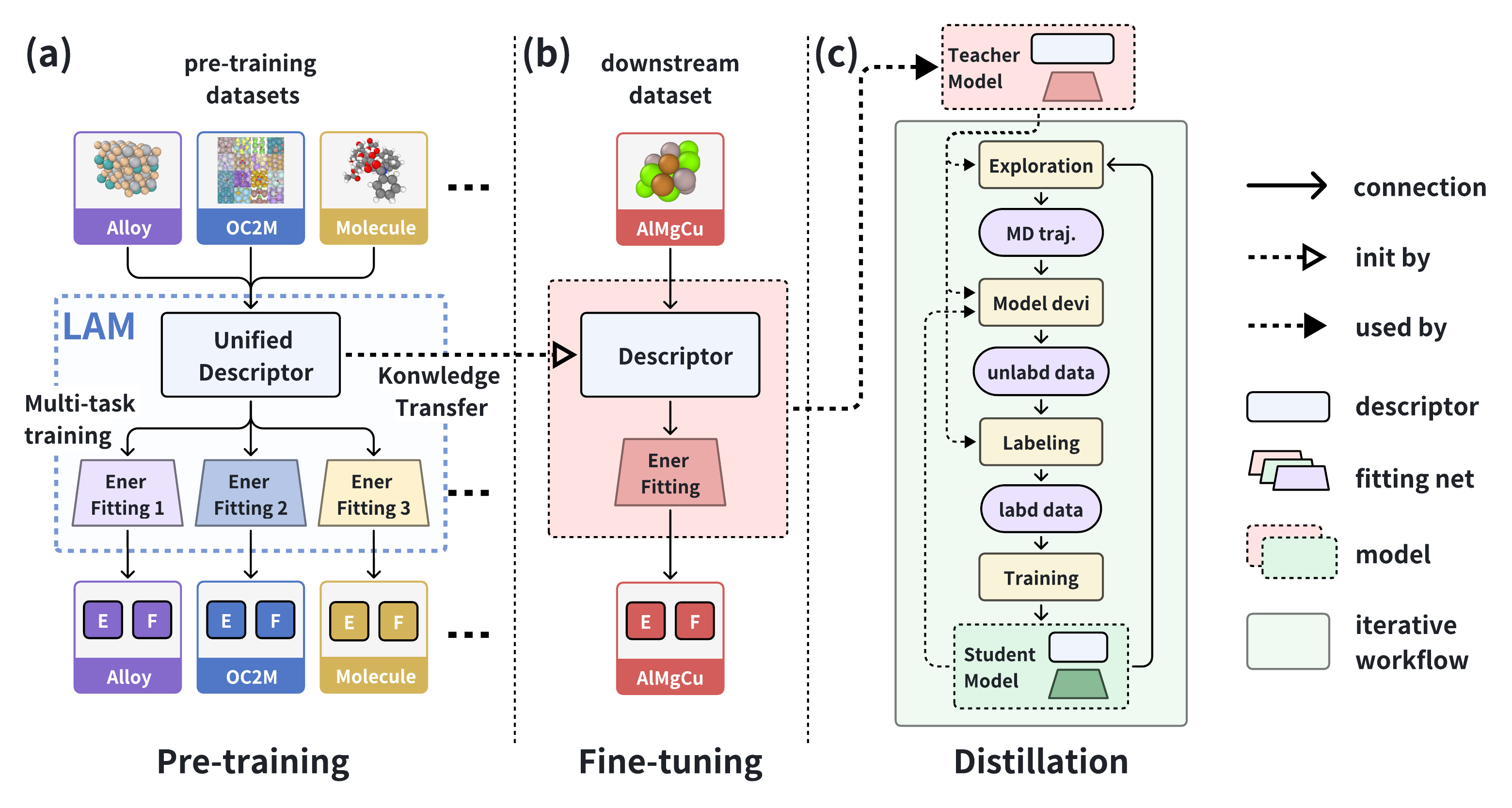}
    \centering
    \caption{An overview of the proposed LAM workflow, (a) the multi-task pre-training process, in which different DFT-labeled data can be pre-trained together by sharing a single descriptor and having their unique fitting nets, with sampling according to their importance. This results in a unified descriptor. (b) The fine-tuning process on the downstream dataset, using the pre-trained unified descriptor and selecting a fitting net from upstream tasks or reinitializing the fitting net for the downstream dataset. (c) The distillation process uses the fine-tuned model as a teacher model, iteratively performing MD simulations and adding labeled data to the training set to train a high-efficiency student model, which is convenient for downstream applications.
    }
    \label{fig:pipeline}
\end{figure*}

\subsection{Related work}

%\WH{MLP architecture. short review. descriptor, GNN, scalability}
\paragraph{Machine learning potential models}
In recent years, there has been rapid development in MLP models. 
While it is nearly impossible to provide a comprehensive list, some notable examples include the Behler-Parrinello neural network (BPNN)~\cite{behler2007generalized}, ANI~\cite{smith2017ani}, deep tensor neural networks (DTNN)~\cite{schutt2017quantum}, weighted atom-centered symmetry functions (wACSF)~\cite{gastegger2018wacsf}, Deep Potential (DP)~\cite{zhang2018deep,zhang2018end,zeng2023deepmd}, Deep Potential with attention (DPA-1)~\cite{zhang2022dpa}, and embedded atom neural network (EANN)~\cite{zhang2019embedded}. 
These models employ either hand-crafted or machine-learned descriptors of atomic environments, along with deep neural networks, to approximate potential energy.
Other machine learning techniques, such as kernel ridge regression, are also widely used. Examples include the Gaussian approximation potential (GAP)~\cite{bartok2010gaussian}, which uses a smooth overlap of atomic positions (SOAP) measure of distance between local environments~\cite{bartok2013representing}, the Coulomb matrix~\cite{rupp2012fast}, and gradient-domain machine learning (GDML)~\cite{chmiela2017machine}. 
Some potential energy models, such as the spectral neighbor analysis method (SNAP)~\cite{thompson2015spectral} and the moment tensor potential (MTP)~\cite{shapeev2016moment}, utilize linear regression for fitting the potential energy surface (PES).

Recently, there has been a surge in the development of equivariant graph neural networks (GNN)~\cite{thomas2018tensor,batzner2021se}, with examples including
SchNet~\cite{schutt2017schnet}, 
Directional Message Passing Neural Network (DimeNet)~\cite{gasteiger2020fast}, 
Polarizable Atom Interaction Neural Network (PaiNN)~\cite{schutt2021equivariant}, 
Geometric Message Passing Neural Network (GemNet)~\cite{gasteiger2021gemnet}, 
SpinConv~\cite{shuaibi2021rotation},
Spherical Channel Network (SCN)~\cite{zitnick2022spherical},
Neural Equivariant Interatomic Potentials (NequIP)~\cite{batzner20223},
MACE~\cite{batatia2022mace}
and Equiformer/EquiformerV2~\cite{liao2022equiformer, liao2023equiformerv2}.
These networks are based on message passing among node and edge equivariant representations and have demonstrated promising fitting accuracy. 
However, it has been noted that GNNs are not easily parallelizable, making them less ideal for large-scale molecular dynamics (MD) simulations~\cite{musaelian2022learning}.

\paragraph{Pre-trained models for molecular modeling} 
Pre-training, or representation learning~\cite{bengio2013representation,zhang2018network}, has shown significant success across various applications, including natural language processing~\cite{devlin2018bert,radford2018improving} and computer vision~\cite{dosovitskiy2020image}. 
In the realm of molecular modeling, a primary objective of pre-trained models is to learn atomic representations of chemical species and 3D configurations of atoms.

One category of downstream tasks involves property prediction.
Pre-trained models can be trained in an unsupervised manner by recovering masked atomic types and perturbed coordinates~\cite{zhou2022uni,zhu2022unified,zaidi2022pretraining,lu2023highly,feng2023fractional}, by undertaking generative tasks~\cite{zhu2022unified}, or by engaging in supervised learning tasks such as regression and classification~\cite{jiao20223d,beaini2023towards,lee2023towards,shoghi2023molecules}.

Another category of downstream tasks focuses on the modeling of PESs.
The model can be pre-trained through unsupervised tasks like denoising or chemical species restoration~\cite{wang2023denoise, cui2023gpip}, supervised learning of energy, force, or partial charge~\cite{gardner2023synthetic,jacobson2023leveraging}, or a combination of both types of tasks~\cite{feng2023may}.
Interestingly, most of these methods were developed for pre-training on molecule-in-vacuum systems, thus limiting the downstream tasks to such a class of tasks.
Ref.~\cite{gardner2023synthetic} developed pre-trained models for condensed-phase carbon systems, but these models are unlikely to be generalizable to systems composed of chemical elements other than carbon.
Zhang et al.~\cite{zhang2022dpa} pre-trained the DPA-1 model on the OC2M dataset~\cite{chanussot2021open} and examined its performance on downstream tasks involving high entropy alloys and AlMgCu ternary alloys. 
However, the study did not investigate downstream tasks related to non-metallic systems.

\section{Results}
\label{sec:results}

\subsection{The workflow of LAM}

% \WH{multi-task training scheme}
The LAM workflow includes the phases of \emph{pre-training}, \emph{fine-tuning} for downstream tasks, and \emph{knowledge distillation}, as schematically presented in Fig.~\ref{fig:pipeline}. The LAM is constructed with a unified descriptor that encodes the symmetry-preserving representation of the chemical and configurational spaces of atomic systems. This descriptor is connected to the energy-fitting networks, each predicting the energy ($E$) and force ($F$) outputs based on the data used during the pre-training phase (see Fig.~\ref{fig:pipeline}(a)).

The LAM employs a multi-task training strategy, as illustrated in Figure~\ref{fig:pipeline}(a). Specifically, the network parameters within the unified descriptor are concurrently optimized through back-propagation using all pre-training datasets. In contrast, the parameters of the fitting network are updated exclusively with the specific pre-training dataset to which they are associated.
This approach is fundamentally different from the single-task training paradigm, where all model parameters, encompassing those within both the descriptor and the fitting network, are refined using a singular training dataset.
The inability to merge the pre-training datasets into a unified ``super-dataset'' stems from the fact that labels across different datasets are typically derived from DFT calculations subject to variable conditions, such as exchange-correlation functionals, basis sets, and energy cut-off radii, culminating in distinct PESs.
We have shown that the multi-task training is as efficient as the single-task training scheme, see Sec.~\ref{sm:sec:multi_task} of the Supplementary Materials.
Therefore, the multi-task training delivers the possibility of training the atomic representation from the heterogeneously labeled pre-training datasets. 
\wh{It is noted that although a hybrid multi-task pre-training approach using both labeled and unlabeled data is technically feasible, we focus on supervised learning for pre-training in this work, and leave the investigation of hybrid multi-task pre-training in future studies.}

% \WH{fine-tuning}
The pre-trained descriptor and the fitting networks can be fine-tuned for specific downstream PES modeling tasks, as illustrated in Figure~\ref{fig:pipeline}(b).
In the downstream model, the descriptor is initialized with the pre-trained unified descriptor, while the fitting network may be initialized either randomly or with a fitting head akin to the one used in one of the pre-training tasks.
Given that the pre-training dataset encodes the bulk of the information within the descriptor, the initialization method for the downstream fitting network is likely to be of minor importance.
The training dataset for a downstream task might be pre-existing and ready for training, or it could be generated through concurrent learning schemes such as DP-GEN~\cite{zhang2020dp}.
In this study, we present several ready-to-use downstream datasets to validate the effectiveness of our proposed methodology and defer the exploration of concurrent learning-based data generation to future research.

% \WH{distillation}
The fine-tuned model, while possessing a large number of parameters, may exhibit reduced efficiency when directly applied to applications like molecular dynamics (MD) simulations. To address this concern, we propose model distillation to create a streamlined version that retains the desired accuracy for downstream tasks while also enhancing processing speed and facilitating extensive simulations.
Figure~\ref{fig:pipeline}(c) depicts the distillation procedure, which employs an iterative learning loop. Within this framework, the original model, henceforth referred to as the ``teacher'', labels the data. In parallel, a ``student'' model, characterized by a simplified architecture (e.g. DPA-1 without any attention layer, which can be further compressed~\cite{lu2022dp} to significantly enhance performance), is trained on this labeled data. 
The teacher model is then engaged in MD exploration, operating under conditions akin to those of the intended downstream application. This ensures that the chemical and physical parameters encountered during both the distillation process and the actual tasks are consistent, facilitating effective learning by the student model. Configurations from the MD trajectories are sampled, and the student model's predictions are compared against those of the teacher. If the discrepancy between their predictions surpasses a pre-established threshold, these configurations are appended to the training set for subsequent iterations.
The cycle is reiterated until the student model's predictive accuracy either meets the preset standards or stabilizes without further improvement.

\subsection{Datasets and DPA-2 descriptor}

% \WH{proposed pre-training datasets}
%The aim of developing LAMs is to encode as much knowledge in the pre-training dataset as possible in the pre-trained model, thus the knowledge would be able to help reduce the effort of fine-tuning the model for a specific downstream task. 
%This gives rise to two requirements in the pre-training stage: 
The primary goal in developing LAMs is to embed comprehensive knowledge within the multi-task pre-trained model by leveraging the pre-training dataset. Consequently, this embedded knowledge is anticipated to alleviate the intensive fine-tuning process required for specific downstream tasks. This objective necessitates two essential criteria during the pre-training phase:
(1) %the pre-training dataset should be able to cover as widely as possible the chemical and configurational spaces that are possible to be encountered in the downstream applications and 
the pre-training dataset must encompass a broad spectrum of chemical and configurational spaces to prepare the model for potential scenarios in downstream applications; and
(2) the DPA-2 model, pre-trained in a multi-task manner, %should have a high capability to generalize to the downstream tasks if the relevant chemical and configurational space of the downstream task is at least partially within the coverage of the pre-training datasets. 
 is expected to exhibit a strong ability to generalize to downstream tasks, provided that the chemical and configurational space relevant to these tasks overlaps to some extent with the scope of the datasets used during pre-training.

\begin{table}
  \caption{%The pre-training and downstream datasets.
  %From left to right columns, we provide the name of the dataset, chemical space coverage, number of training data, number of validation data, the total number of data, and the weight in the pre-training. 
  Overview of pre-training and downstream datasets employed in the multi-task learning framework. The columns provide dataset name, coverage of the chemical space, number of training data points, number of test data points, the total data count, and assigned weight.
  }
  \label{tab:dataset}
  \footnotesize
  \centering
    \begin{tabular}{llR{1.5cm}R{1.5cm}R{1.5cm}R{1.0cm}}
    \toprule
    \multicolumn{6}{c}{Pre-training datasets} \\
    Name & element & \#train & \#test & \#total & weight \\
    \hline
    Alloy               & 53                        &     71,482   &      1,240   &     72,722   &   2.0 \\
    Cathode-P           & Li,Na,O,Mn,Fe,Co,Cr,Ni    &     58,690   &      6,451   &     65,141   &   1.0 \\
    Cluster-P           & Pd,Ru,Al,Au,Ag,Pt,Si,Cu,Ni&    139,200   &     14,936   &    154,136   &   1.0 \\
    Drug                & H, C, N, O, F, Cl, S, P   &  1,379,956   &     24,257   &  1,404,213   &   2.0 \\
    FerroEle-P          & 15                        &      6,966   &        760   &      7,726   &   1.0 \\
    OC2M                & 56                        &  2,000,000   &    999,866   &  2,999,866   &   2.0 \\
    SSE-PBE-P           & Li, P, S, Si, Ge          &     15,019   &        755   &     15,774   &   1.0 \\
    SemiCond-P          & 14                        &    136,867   &     14,848   &    151,715   &   1.0 \\
    H2O-PD              & H, O                      &     46,077   &      2,342   &     48,419   &   1.0 \\
    Ag$\cup$Au-PBE      & Ag, Au                    &     16,696   &        812   &     17,508   &   0.2 \\
    Al$\cup$Mg$\cup$Cu  & Al, Mg, Cu                &     24,252   &      1,145   &     25,397   &   0.3 \\
    Cu                  & Cu                        &     14,596   &        770   &     15,366   &   0.1 \\
    Sn                  & Sn                        &      6,449   &        276   &      6,725   &   0.1 \\
    Ti                  & Ti                        &     10,054   &        474   &     10,528   &   0.1 \\
    V                   & V                         &     14,935   &        738   &     15,673   &   0.1 \\
    W                   & W                         &     42,297   &      2,100   &     44,397   &   0.1 \\
    C12H26              & H, C                      &     33,898   &      1,598   &     35,496   &   0.1 \\
    HfO2                & O, Hf                     &     27,660   &        917   &     28,577   &   0.1 \\
    sum                 & 73                        &  4,045,094   &  1,074,285   &  5,119,379   &  13.2 \\
    \hline
    \multicolumn{6}{c}{Downstream datasets} \\
    Name            & element & \#train & \#test & \#total & weight \\
    \hline
    Cathode-D           & Li, Na, O, Mn, Fe, Co, Cr      &     30,002   &      3,244   &     33,246    & 1.0 \\
    Cluster-D           & Pd, Au, Ag, Pt, Cu, Ni         &      4,218   &        395   &      4,613    & 1.0 \\
    FerroEle-D          & 15                             &      7,521   &        597   &      8,118    & 1.0 \\
    SSE-PBE-D           &  Li, P, S, Sn                  &      2,563   &        131   &      2,694    & 0.5 \\
    SSE-PBESol          &  Li, P, S, Si, Ge, Sn          &      7,502   &        384   &      7,886    & 0.5 \\
    SemiCond-D          &P, N, Al, Te, In, Se, Sb, B, As &     78,614   &      8,495   &     87,109    & 1.0 \\
    ANI-1x              &  H, C, N, O                    &  4,872,049   &     83,956   &  4,956,005    & 1.0 \\
    Transition-1x       &  H, C, N, O                    &  7,632,328   &    967,454   &  8,599,782    & 1.0 \\
    H2O-DPLR            &  H, O                          &        557   &         46   &        603    & 0.5 \\
    H2O-SCAN0           &  H, O                          &      7,002   &        347   &      7,349    & 0.5 \\
    H2O-PBE0TS          &  H, O                          &    133,000   &      7,000   &    140,000    & 0.5 \\
    H2O-PBE0TS-MD       &  H, O                          &     38,000   &      2,000   &     40,000    & 0.5 \\
    AgAu-PBED3          &  Ag, Au                        &     64,239   &      2,256   &     66,495    & 0.3 \\
    AlMgCu-D            &  Al, Mg, Cu                    &    113,942   &      2,820   &    116,762    & 0.2 \\
    In2Se3              &  In, Se                        &     11,621   &        568   &     12,189    & 0.2 \\
    sum                 &  39                            & 13,003,158   &  1,079,693   & 14,082,851    & 9.0 \\
    \toprule            
  \end{tabular}
\end{table}

For the first criterion, the datasets utilized in this study are summarized in Table~\ref{tab:dataset}. Detailed descriptions are provided in Section~\ref{sm:sec:data} of the Supplementary Materials.
Some datasets are newly generated in this work, including metallic alloys (Alloy), cathode materials (Cathode), metal nano-clusters (Cluster), and drug-like molecules (Drug). 
Some datasets are contributed by the DeepModeling community~\footnote{See \url{https://github.com/deepmodeling/AIS-Square/tree/main/datasets}}, including the ferroelectric perovskite (FerroEle), solid-state-electrolyte (SSE), semiconductors (SemiCond), \ce{H2O}, metallic material datasets (e.g.~Sn, AgAu and AlMgCu), and the pyrolysis of n-dodecane (C12H26). 
Additionally, we have the open catalyst 20~\cite{chanussot2021open} (OC2M) that is formed by AIMD trajectories of molecular chemical reactions catalyzed by metallic substrates. 
These datasets are labeled with various DFT software like the VASP~\cite{kresse1996efficiency,kresse1996efficient}, Gaussian~\cite{frisch2016gaussian}, and ABACUS~\cite{chen2010systematically, li2016large}.  
In addition, They are divided into two groups, the pre-training and the downstream datasets, as detailed in Sec.~\ref{sm:sec:data} of the Supplementary Materials. 
It is noted that the division is only to demonstrate the effectiveness of the workflow of LAM. 
For production purposes, all the datasets listed in Tab.~\ref{tab:dataset} should be used to pre-train a LAM. 

In the last column of Tab.~\ref{tab:dataset}, weights are assigned to each pre-training dataset. These weights are based on relevance, diversity in both chemical and configurational spaces, and data volume. The weight of a dataset is proportional to its selection probability during multi-task training, meaning that datasets with higher weights are favored in each training iteration. 
These weights also play a crucial role in calculating the weighted average of errors across all datasets, as shown in Tab.~\ref{tab:zero-shot} and Tabs.~\ref{sm:tab:single_task}--\ref{sm:tab:multi_task} of the Supplementary Materials, which helps to provide an assessment of the model's overall accuracy.

\begin{figure*}
    \includegraphics[width=1.0\textwidth]{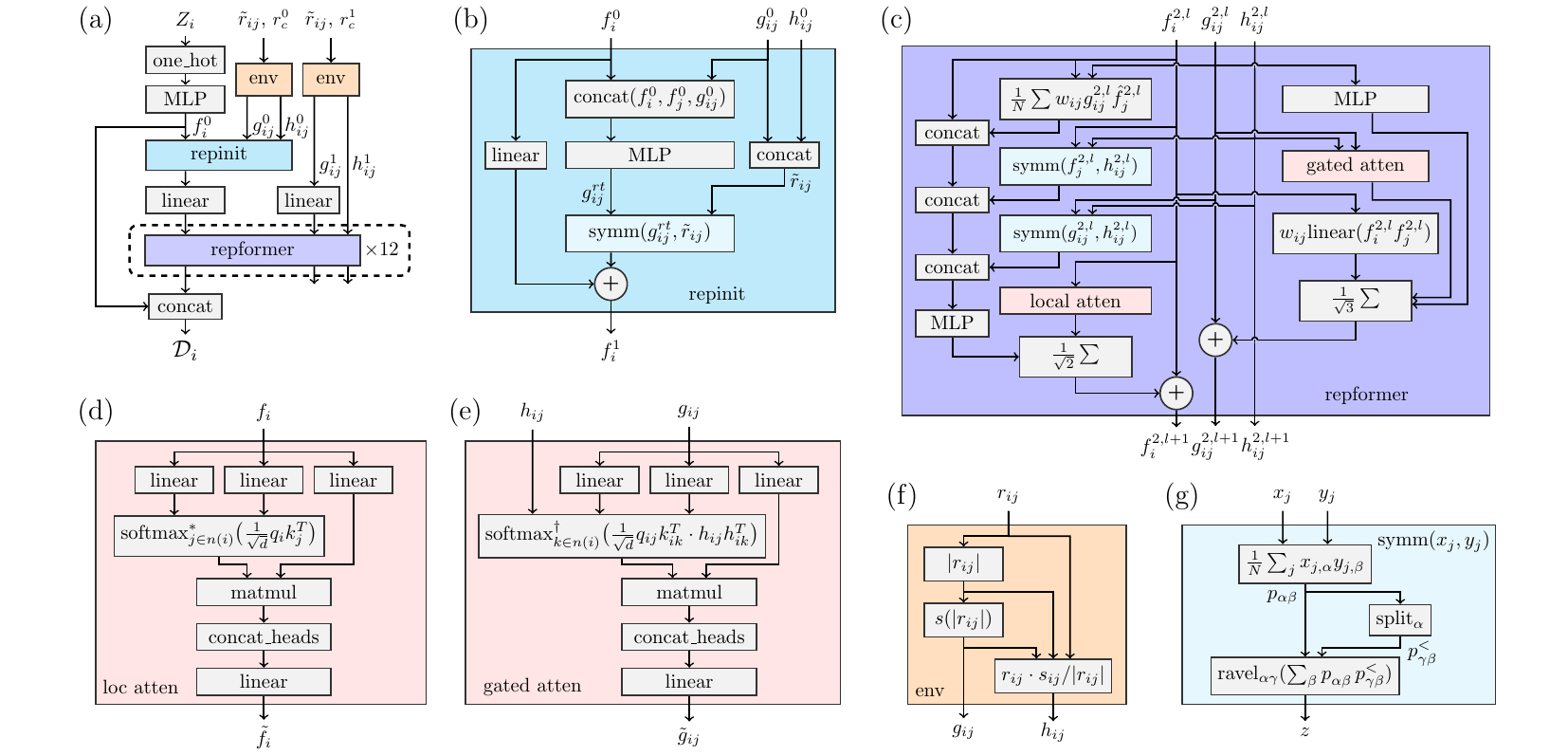}
    \centering
    \caption{(a) Detailed architecture of the DPA-2 descriptor, which includes two primary components: $\mathrm{repinit}$ and $\mathrm{repformer}$. (b) Structure of $\mathrm{repinit}$. (c) Structure of $\mathrm{repformer}$. (d-g) Substructures referenced in subsequent sections.
    }
    \label{fig:model}
\end{figure*}

For the second criterion, we propose the DPA-2 model with full details of the model architecture explained in Section~\ref{sec:method}.
The descriptor of the model, which is supposed to encode the representation of the chemical and configurational spaces of the pre-training dataset, is schematically demonstrated in Fig.~\ref{fig:model}. 
The chemical and configurational spaces are represented by a single-atom channel $f_i$, a rotationally invariant pair-atom channel $g_{ij}$ and a rotationally equivariant pair-atom channel $h_{ij}$.
The pair-atom representations are initialized by the environment matrix (operator $\mathrm{env}$ in Fig.~\ref{fig:model}), which encodes the relative positions of the near neighbors within a certain cut-off radius ($r_c^0$ and $r_c^1$), and smoothly decays to zero at the cut-off radius. 
The single-atom representations $f_i$ is initialized by a $\mathrm{repinit}$ (representation initializer) layer.
Then the single- and pair-atom representations are subsequently updated by the representation transformer ($\mathrm{repformer}$) layers, which are stacked 12 times and communicate information in a message-passing manner between the layers.  
In each of the $\mathrm{repformer}$ layer, %single-atom representation 
$f_i$ is updated by convolution, symmetrization, MLP, and localized self-attention operators, while %the rotationally invariant representation 
$g_{ij}$ is updated by MLP, dot-product, and gated self-attention operators (see Fig.~\ref{fig:model}(c) and Sec.~\ref{sec:dpa2} for more details). 
% The updating of rotationally equivariant representation $h_{ij}^0$ leads to a decrement in the stability of the model, thus is not included in the $\mathrm{repformer}$ layer. 
The contribution of different building blocks to the model accuracy is investigated by an ablation study in Sec.~\ref{sm:sec:ablation} of the Supplementary Materials. 

The DPA-2 model is designed to be extensible and inherently respects translational, rotational, and permutational symmetries.
%The DPA-2 model is conservative, because it predicts the atomic forces by the negative gradients of the system energy with respect to the atomic positions, $F_i = -\nabla_{r_i} E$, and calculates the virial tensor by $\Xi_{\alpha\beta} =  \sum_\gamma (-\nabla_{h_{\gamma\alpha}}E) h_{\gamma\beta}$, where $E$ being the energy, $r_i$ being the position of atom $i$ and $h_{\alpha\beta}$ being the $\beta$th component of the $\alpha$th basis vector of the simulation cell. 
Moreover, it is conservative, as it predicts atomic forces by computing the negative gradient of the system's energy with respect to the atomic positions, $F_i = - \nabla_{r_i} E$, and calculates the virial tensor as $\Xi_{\alpha\beta} =  \sum_\gamma (-\nabla_{h_{\gamma\alpha}}E) h_{\gamma\beta}$, where $E$ represents the energy, $r_i$ denotes the position of atom $i$, and $h_{\alpha\beta}$ is the $\beta$th component of the $\alpha$th basis vector of the simulation cell.
Furthermore, all components of the DPA-2 model are continuous up to the second-order derivative, ensuring energy conservation.
Numerical examples demonstrating the energy conservation properties of the DPA-2 model can be found in Supplementary Material Sec.~\ref{sm:sec:conserv}.

\subsection{Generalizability of the multi-task pre-trained DPA-2 model}

\begin{table}
% \scriptsize
  \caption{
    Comparison on the zero-shot generalization errors on downstream tasks.
    The MACE-MP-0 (MACE) and DPA-2 pre-trained on MPtrj dataset, the DPA-2 pre-trained by single-task (ST) and multi-task (MT) approaches are compared.
  The DPA-2 ST is trained by the pre-training datasets listed in the second column of the Table, 
  while the DPA-2 MT is trained by all the pre-training datasets listed in Tab.~\ref{tab:dataset}.
  The energy and force RMSEs on the downstream test datasets are reported.
  The weighted averaged RMSEs (WARMSE) with the weights presented in Tab.~\ref{tab:dataset} is given in the first row of the table.
  The standard deviations of energy and force labels in the test set are also provided.
  If the RMSE is smaller than the corresponding standard deviation, the model shows the ability of zero-shot generalization, on the other hand, the model cannot be generalized to downstream tasks without downstream data. 
  }
  \scriptsize
  % Zero-shot generalization errors of the models trained by single-task (ST) or multi-task (MT) approach.
  % The ST models are trained by the pre-training datasets listed in the left most column of the Table, 
  % while the DPA-2 MT is trained by all the pre-training datasets listed in Tab.~\ref{tab:dataset}.
  % The ST-trained four models are compared, the GNO (GemNet-OC), EFV2 (EquiformerV2), Allegro and DPA-2.
  % The energy and force RMSEs on the downstream validation datasets are reported.
  % The standard deviations of energy and force labels in the validation set are also provided.
  % If the RMSE is smaller than the corresponding standard deviation, the model show ability of zero-shot generalization, on the other hand, the model cannot be generalized to downstream tasks without downstream data. 
  % The ``--'' in the table means the pre-training failed with CUDA out-of-memory error.
  \setlength{\tabcolsep}{4pt}
  \label{tab:zero-shot}
  \centering 
    \begin{tabular}{ll| R{0.8cm}R{0.8cm}R{0.8cm}R{0.8cm}R{0.8cm} |  R{0.8cm}R{0.8cm}R{0.8cm}R{0.8cm}R{0.8cm}  }
      \toprule
               &                & \multicolumn{5}{c|}{Energy RMSE [meV/atom] $\downarrow$} &\multicolumn{5}{c}{Force RMSE [meV/\AA] $\downarrow$} \\\cline{3-12}
Downstream     &Pre-train       &data   &MACE   &DPA-2  &DPA-2  &DPA-2  &data   &MACE   &DPA-2  &DPA-2  &DPA-2  \\
               &(only for ST)   &std.   &(MPtrj)&(MPtrj)&ST     &MT     &std.   &(mptrj)&(mptrj)&ST     &MT     \\\hline
\bf{WARMSE}    &     &\bf{121.4	}&\bf{104.0	}&\bf{68.3	}&\bf{100.2	}&\bf{50.1	}&\bf{1405.4	}&\bf{575.6	}&\bf{516.6	}&\bf{628.0	}&\bf{238.8}\\
AgAu-PBED3     & AgAu-PBE	&906.9	&1812.8	&268.9	&222.9	&192.3	&878.0	&683.2	&293.3	&236.9	&63.6   \\
AlMgCu-D       & AlMgCu         &383.8	&33.8	&32.0	&254.3	&41.2	&1229.5	&240.1	&245.3	&663.7	&111.8  \\
AlMgCu-D       & Alloy          &383.8	&33.8	&32.0	&74.9	&48.4	&1229.5	&240.1	&245.3	&122.3	&112.8  \\
ANI-1x         & Drug           &198.9	&52.3	&61.7	&67.2	&56.6	&2124.6	&636.1	&700.1	&738.7	&346.7  \\
Cathode-D      & Cathode-P	&42.2	&15.8	&29.7	&39.8	&43.8	&641.9	&288.4	&613.9	&339.7	&273.9  \\
Cluster-D      & Cluster-P	&636.0	&323.7	&262.7	&41.4	&40.5	&3605.4	&2230.8	&1193.6	&238.4	&190.5  \\
FerroEle-D     & FerroEle-P	&43.0	&12.5	&14.5	&6.3	&3.9	&881.3	&191.3	&194.2	&282.7	&115.1  \\
H2O-DPLR       & H2O-PD         &15.6	&2.1	&2.0	&9.1	&9.3	&825.2	&94.4	&99.7	&263.5	&263.4  \\
H2O-H2O        & H2O-PD         &47.0	&4.9	&7.2	&4.9	&4.7	&1941.0	&381.0	&382.7	&58.8	&64.4   \\
H2O-PBE0TS-MD  & H2O-PD         &3.3	&1.1	&1.5	&0.5	&0.6	&816.1	&330.8	&314.4	&37.6	&40.8   \\
H2O-SCAN0      & H2O-PD         &12.6	&3.2	&3.8	&1.1	&0.7	&2163.2	&387.5	&385.2	&409.2	&162.9  \\
In2Se3         & SemiCond-P	&120.5	&31.9	&24.5	&160.6	&38.9	&611.1	&190.2	&188.0	&1544.1	&341.6  \\
SemiCond-D     & SemiCond-P	&587.6	&49.8	&70.9	&486.2	&175.7	&1755.4	&470.7	&534.9	&1439.4	&439.3  \\
SSE-PBE-D      & SSE-PBE-P	&79.0	&33.7	&39.4	&40.7	&6.2	&789.5	&222.1	&249.9	&635.6	&162.4  \\
SSE-PBESol     & SSE-PBE-P	&84.3	&32.5	&37.4	&26.1	&8.3	&810.9	&231.8	&260.4	&425.0	&115.3  \\
Transition-1x  & Drug           &139.8	&56.4	&55.1	&48.2	&45.8	&368.1	&518.6	&618.3	&1298.6	&363.8  \\
    \bottomrule    
  \end{tabular}
\end{table}

Before moving to a discussion on the generalizability of the multi-task training scheme, we test the model of DPA-2 by using single-task benchmarks, which are directly comparable to the state-of-the-art model architectures. 
In the first benchmark, the ANI-1x dataset, the DPA-2 shows superior test accuracy compared with the ANI-1x model reported in Ref.~\cite{smith2018less}, see Tab.~\ref{sm:tab:ani-test} in the Supplementary Materials. 
In the second benchmark, the accuracy of the DPA-2 model is comparable to GemNet-OC~\cite{gasteiger2022gemnet} and higher than Equiformer V2~\cite{liao2023equiformerv2}, NequIP~\cite{batzner20223}, Allegro~\cite{musaelian2022learning} and MACE~\cite{batatia2022mace} models on the pre-training datasets, see Tab.~\ref{sm:tab:single_task} in the Supplementary Materials.

% To compare the generalizability with ML PES models, the DPA-2 model is firstly trained by the single-task scheme on the ANI-1x and all the pre-training datasets. 
% On the ANI-1x dataset, the DPA-2 shows superior test accuracy compared with the ANI-1x model reported in Ref.~\cite{smith2018less}, see Tab.~\ref{sm:tab:ani-test} in the Supplementary Materials. 
% On the pre-training datasets, the accuracy of the DPA-2 model is comparable to GemNet-OC~\cite{gasteiger2022gemnet} and higher than Equiformer V2~\cite{liao2023equiformerv2}, NequIP~\cite{batzner20223}, and Allegro~\cite{musaelian2022learning} models, see Tab.~S2 in the Supplementary Materials. 
% In this benchmark, the test data are of the same distribution as the training data, thus a higher accuracy does not necessarily indicate a better performance in the pre-training-and-fine-tuning scenario. 

Next, we train the DPA-2 model on all the pre-training datasets by the multi-task scheme. 
The details of the training protocol, the test accuracy of these datasets, and a discussion on the effectiveness of the multi-task scheme are given in Sec.~\ref{sm:sec:multi_task} of the Supplementary Materials.  
% \LZ{give links and section name?}

%The generalizability of the multi-task pre-trained DPA-2 model to downstream tasks is investigated by directly test the model on the downstream datasets. 
%Since no datum from the downstream is used to improve the quality of the pre-trained model before testing, such a generalization is called the zero-shot generalization. 
%If the pre-trained model were perfectly generalizable, i.e.~were able to represent the chemical knowledge of the periodic table and all the relevant configurations of a downstream task, the zero-shot generalization error would be the same as, if not even lower than, the validation error of a  model properly trained for that specific downstream task from scratch.
%Usually some degree of the zero-shot generalizability presents, in the sense that the RMSE errors are observed to be smaller than the standard deviation of the labels, but there still exists a gap from the perfect generalization. 
%In this scenario, we need to fine-tune the pre-trained model by data from the downstream task.
%If a pre-trained model processes a stronger generalizabilty, the less data would be needed in the fine-tuning, and the amount of saved data in comparison with a from-scratch training measures the benefit we would expect from using a pre-trained model. 
We investigate the generalizability of the multi-task pre-trained DPA-2 model to downstream tasks by testing the model directly on downstream datasets. This approach is known as zero-shot generalization because no data from the downstream tasks are used to refine the pre-trained model before testing. In an ideal scenario, a perfectly generalizable model—that is, one that encapsulates the chemical knowledge of the periodic table and all relevant configurations for a given downstream task—would exhibit a zero-shot generalization error comparable to, or potentially lower than, the test error of a model specifically trained from scratch for that task.

The zero-shot generalizability of the multi-task pre-trained DPA-2 model is presented in Table~\ref{tab:zero-shot} and compared with its single-task pre-trained counterpart, MPtrj-trained DPA-2, and MACE-MP-0.
For all cases, the single-task DPA-2 models are exclusively trained on the datasets specified in the second column, whereas the multi-task DPA-2 model undergoes pre-training on the entire corpus of pre-training datasets (see Table~\ref{tab:dataset}).
The multi-task DPA-2 model then employs the fitting head indicated in the second column to initialize the fitting procedure for downstream tasks.
All model variants are evaluated on their respective downstream datasets without any additional training.
The results demonstrate that multi-task training substantially enhances generalizability compared to the single-task pre-trained DPA-2 and the MPtrj-trained models.
The comparable performance between the MPtrj-trained MACE-MP-0 and DPA-2 suggests that the improvement is primarily due to the multi-task pre-training scheme rather than differences in model architecture.

\subsection{Fine-tuning on downstream tasks}

\begin{figure*}
    \includegraphics[width=1.0\textwidth]{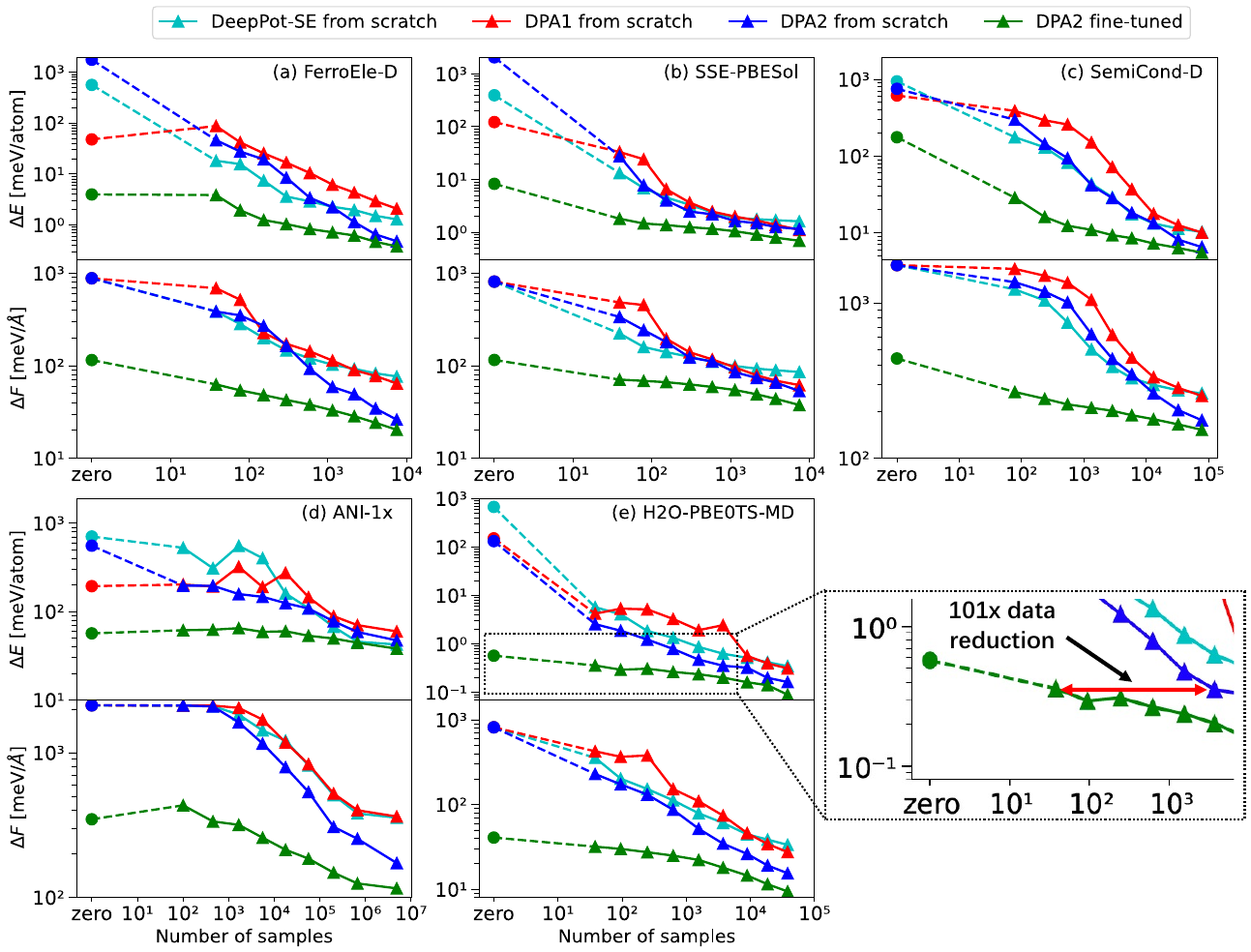}
    \centering
    \caption{Comparative analysis of sample efficiency on downstream tasks. The horizontal axis represents the volume of downstream data required, while the vertical axis depicts the RMSE convergence in energy or force predictions. For a uniform assessment across models, the number of training epochs per model for each downstream task is normalized to a standard value, derived by dividing 1 million by the number of downstream samples.
    }
    \label{fig:lcurve_small}
\end{figure*}

%The performance of the fine-tuned DPA-2 model in downstream tasks is investigated by comparing the sample efficiency with several DP models trained from scratch. 
%In Figure~\ref{fig:lcurve_small}, several representative downstream tasks are presented, while a complete comparison can be found in Sec.~S4 of the Supplementary Materials. 
%In the Figure, the trends in the convergence of the energy and force RMSEs are plotted against an increasing volume of downstream data.

Although zero-shot generalizability is often observed to a certain extent, a gap from perfect generalization typically remains. To bridge this gap, we fine-tune the models using data from the downstream tasks.
A stronger generalizability in a pre-trained model implies that less data is required during fine-tuning, leading to higher sample efficiency. The reduction in sample size relative to training a model from scratch quantifies the advantage of employing a multi-task pre-trained model.

The sample efficiency of the pre-trained DPA-2 on downstream tasks was evaluated by comparing it against various other DP models that were trained from scratch.
Fig.~\ref{fig:lcurve_small} showcases a selection of downstream tasks, with a comprehensive comparison available in Section~\ref{sm:sec:lcurve} of the Supplementary Materials.
The figure illustrates the convergence trends of the energy and force RMSEs in relation to the expanding sample size used for downstream training.

To draw distinctions between the fine-tuned DPA-2 and the from-scratch DPA-2 models, it is important to realize that both models share identical architectures. However, the fine-tuned model begins with parameters derived from a multi-task pre-trained model, whereas the from-scratch model starts with randomly initialized parameters. 
The fine-tuned DPA-2 model consistently achieves lower error curves compared to the DPA-2 model trained from scratch, particularly when the available downstream data is scarce.
This translates to a considerable reduction in the amount of data needed to reach equivalent levels of accuracy. 
Taking the H2O-PBE0TS-MD task for example, two orders of magnitudes of training data are saved to reach the same energy accuracy, see the zoomed-in of Fig.~\ref{fig:lcurve_small}. 
As the sample size grows, the performance disparity between the fine-tuned and from-scratch DPA-2 models diminishes. This outcome is anticipated, given that both models possess the same capacity and, theoretically, their accuracy should converge as the dataset approaches an infinite size.
When comparing DeepPot-SE (DP-SE), DPA-1, and DPA-2 models trained from scratch, the DPA-2 model exhibits superior performance over the other architectures. 
While the convergence patterns of the DPA-1 and DP-SE models are somewhat parallel, the DP-SE model reaches a performance plateau more rapidly than the DPA-1 in the FerroEle-D, SSE-PBESol, and SemiCond-D tasks.

\subsection{Model distillation and evaluation}

\begin{figure*}
    \includegraphics[width=1.0\textwidth]{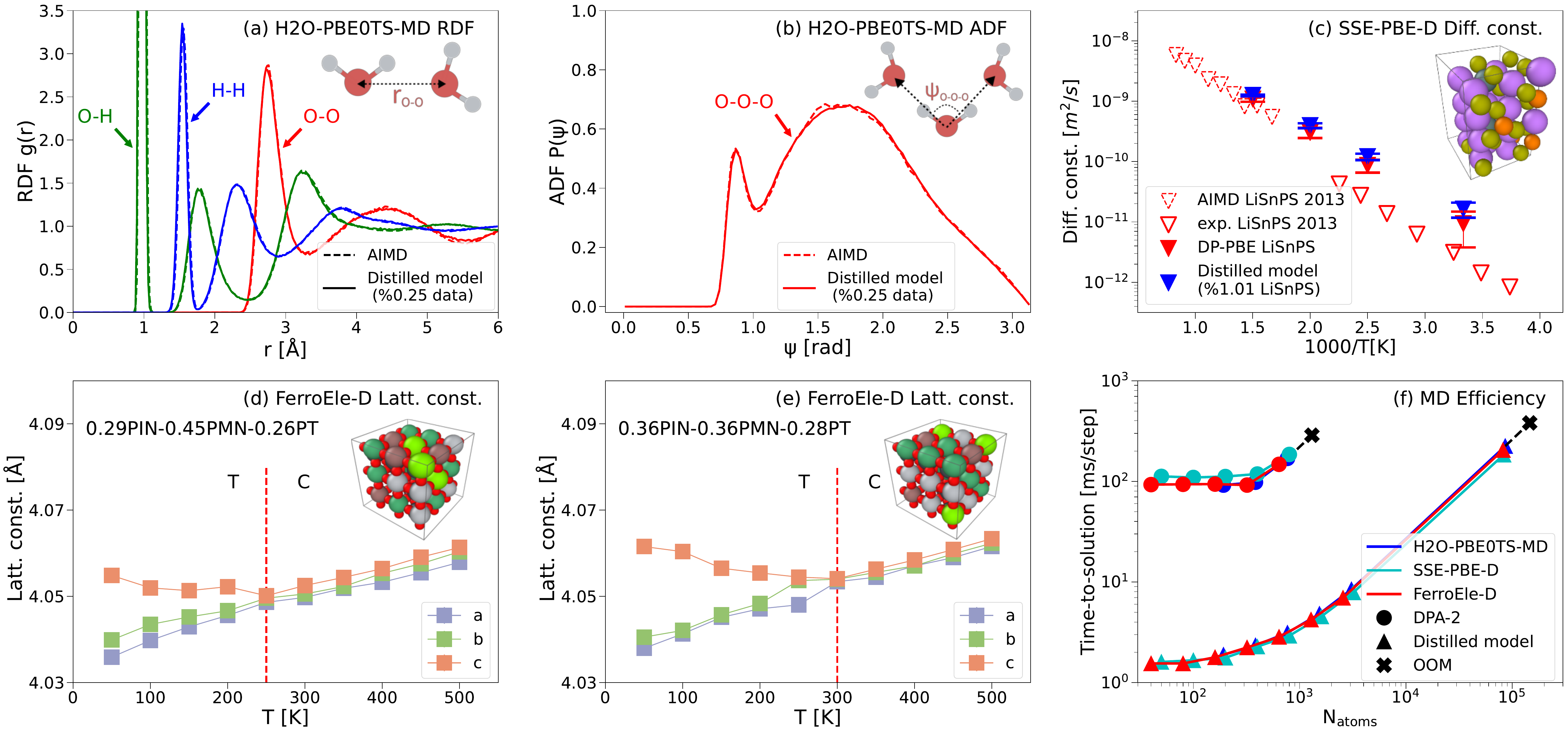}
    \centering
    \caption{Evaluation of the distilled model across various downstream applications.
    (a-b) Comparison of the radial distribution function (RDF) and angular distribution function (ADF) for the H2O-PBE0TS-MD dataset between the reference AIMD results~\cite{distasio2014individual} and the distilled model. 
    The model is distilled from a DPA-2 model fine-tuned from merely 0.25\% of DFT-labeled data.
    (c) A comparison of diffusion constants for the solid-state electrolyte \ce{Li10SnP2S12}. 
    The constants were determined using various methods: the distilled model, DPMD as reported in Huang et al.~(2021)~\cite{huang2021deep}, AIMD simulations from the studies by Mo et al.~(2012) and Marcolongo et al.~(2017)~\cite{mo2012first, marcolongo2017ionic}, and experimental findings from solid-state nuclear magnetic resonance (NMR) as documented by Kuhn et al.~(2013)~\cite{kuhn2013single}.
    The distilled model is trained from a DPA-2 model fine-tuned by 1.01\% of the SSE-PBE-D data.
    (d-e) The temperature-dependent lattice constants for the ternary solid solution ferroelectric perovskite oxides \ce{Pb(In1/2 Nb1/2 )O3}–\ce{Pb(Mg1/3 Nb2/3 )O3}–\ce{PbTiO3} (PIN-PMN-PT). 
    The NPT MD simulations using the distilled model are conducted for two concentrations, 0.29PIN–0.45PMN–0.26PT and 0.36PIN-0.36PMN-0.28PT~\cite{wu2023universal}.
    The model is distilled from a DPA-2 model fine-tuned with the complete FerrEle-P dataset and 7.86\% of the FerrEle-D data. 
    (f) Computational efficiency assessment for the aforementioned three systems, showcasing the time-to-solution as a function of the system size in the number of atoms ($N_{\mathrm{atoms}}$).
    }
    \label{fig:evaluation}
\end{figure*}

%The fine-tuned DPA-2 model is usually computationally inefficient due to the large number of parameters, see Fig.~\ref{fig:evaluation}(f). 
%We thus distill the knowledge of the fine-tuned models to compressed DPA-1 models without attention layers and investigate the efficiency and accuracy of the distilled models on three representative downstream tasks H2O-PBE0TS-MD, SSE-PBE-D, and FerroEle-D. 
The fine-tuned DPA-2 model typically suffers from computational inefficiency due to its extensive parameter set, as illustrated in Fig.~\ref{fig:evaluation}(f). 
To address this, we employed a knowledge distillation approach, transferring insights from the fine-tuned DPA-2 models to compressed DPA-1 models without attention layers. We evaluated the performance of these distilled models in terms of efficiency and accuracy on three benchmark downstream tasks: H2O-PBE0TS-MD, SSE-PBE-D, and FerroEle-D.
Notably, in all the cases, the fine-tuned models are exposed to only a small portion (0.25\%--7.86\%, see Tab.~\ref{sm:tab:table_distill}) of the downstream dataset, and are used to generate the distillation training datasets that sufficiently cover the relevant configuration spaces. 
In the FerroEle-D task, we append the full FerroEle-P to a small (7.86\%) portion of the FerroEle-D dataset for the training of the fine-tuned model.
The FerroEle-D that contains solid solution perovskite oxides was generated by the concurrent learning scheme starting from the FerroEle-P dataset that contains unitary perovskite (see Ref.~\cite{wu2023universal} and Supplementary Materials Sec.~\ref{sm:sec:data}). %Therefore,  the full FerroEle-D dataset itself is not sufficient to train a qualified potential model. 
Consequently, the FerroEle-D dataset alone does not provide a comprehensive basis for training a fully capable potential model.

After distillation, the time-to-solution and the maximal system size that can be simulated on a single GPU card improved by nearly two orders of magnitude, as shown in Fig.~\ref{fig:evaluation}(f).
Moreover, the accuracy of the distilled models is on par with that of the fine-tuned DPA-2 models, as detailed in Tab.~\ref{sm:tab:table_distill}.
The distilled models appear to have reached the peak of their performance, given that their accuracies closely match those of the DPA-1 models (without an attention layer) when trained on the complete downstream datasets.

% After fine-tuning, we obtained a high-precision downstream model. To further enhance the efficiency of this model in practical MD simulations for downstream tasks, we employed model distillation as introduced in Section~\ref{sec:distill}. We conducted model distillation on the H2O-PBE0TS-MD, SSE-PBE-D, and FerrEle-D datasets, the detailed settings and distillation accuracy are shown in Sec.~S6 of the Supplementary Materials. 
% We can observe that after distillation, the student model, which is distilled from the teacher model fine-tuned with a small amount of data, is capable of achieving its ultimate accuracy trained on the full dataset. Concurrently, Figure.~\ref{fig:evaluation}(f) demonstrates the efficiency of MD conducted on these three datasets by models before and after distillation. It is apparent that the post-distillation student model outpaces the teacher model by nearly a hundredfold in efficiency across various system sizes, which is much more suitable for efficient and large-scale simulations.

Finally, to validate the reliability of the distilled models beyond the energy and force RMSEs, we have conducted various application tests on the aforementioned three systems, as reported in Fig.~\ref{fig:evaluation}(a-e).
In the downstream task of H2O-PBE0TS-MD, we observe that the radial distribution functions (RDFs) and the angular distribution function (ADF) of the distilled model are in almost perfect agreement with those obtained from the AIMD simulation, see Fig.~\ref{fig:evaluation}(a-b).
In the downstream task of SSE-PBE-D, the diffusion constants of Lithium ions in the \ce{Li10SnP2S12} system under different temperature conditions are calculated. 
The distilled model presents satisfactory agreement with the previously reported MD simulations using DP-PBE LiSnPS model and DFT (i.e.~AIMD simulations)~\cite{mo2012first, marcolongo2017ionic}, see Fig.~\ref{fig:evaluation}(c). 
The discrepancy between the simulation and the experimental results~\cite{kuhn2013single} may be attributed to the approximation error of the density functional and finite size effects, as discussed in Ref.~\cite{huang2021deep}. 
In the downstream task of FerroEle-D, we investigated the temperature-driven phase transition in the solid solution ferroelectric perovskite \ce{Pb(In1/2 Nb1/2 )O3}–\ce{Pb(Mg1/3 Nb2/3 )O3}–\ce{PbTiO3} (PIN-PMN-PT), see Fig.~\ref{fig:evaluation}(d-e).
Tetragonal-cubic (T-C) transitions are observed at $\sim 250$ and $\sim 300$~K for two concentrations 0.29PIN–0.45PMN–0.26PT and 0.36PIN-0.36PMN-0.28PT, repectively. 
The fact that the transition temperature raises for $\sim 50$~K due to the increment in the PIN (\ce{Pb(In1/2 Nb1/2 )O3}) portion from 29\% to 36\% is in line with the experimental observations~\cite{wang2012phase,li2018soft}.

% Moreover, to validate the reliability of the model after distillation, beyond ensuring accuracy, we have conducted various application tests on the aforementioned three systems, as depicted in Figure~\ref{fig:evaluation}(a-e).
% It should be noted that although the student model is exposed to only a very limited amount of the original data (less than 2\%) during the entire distillation process, the teacher model further explores and labels additional data for the student model to learn from, thereby ensuring coverage of the configurations required for the downstream tasks.
% In the application results, we observe that on the H2O-PBE0TS-MD dataset, the curves of the RDF and the ADF generated by the distilled model are in complete agreement with those obtained from DFT calculations.
% For the diffusion constants on the SSE-PBE-D dataset, the performance of the distilled model is also consistent with experimental data and previous DP-PBE model, though the discrepancy between the models and experimental results may be attributed to finite size effects.
% In the testing of lattice constants as a function of temperature on the FerroEle-D dataset, the distilled model accurately reproduces the phase transition temperature; however, at low temperatures, the model estimates for the c-axis are lower than those reported in~\cite{wu2023universal}, which could be due to a combination of factors such as model training and data.
% Overall, the distilled model achieves reasonable results that are in good agreement with experimental outcomes in its downstream applications.

\subsection{The representation learned by the DPA-2 model}

\begin{figure*}
    \includegraphics[width=1.0\textwidth]{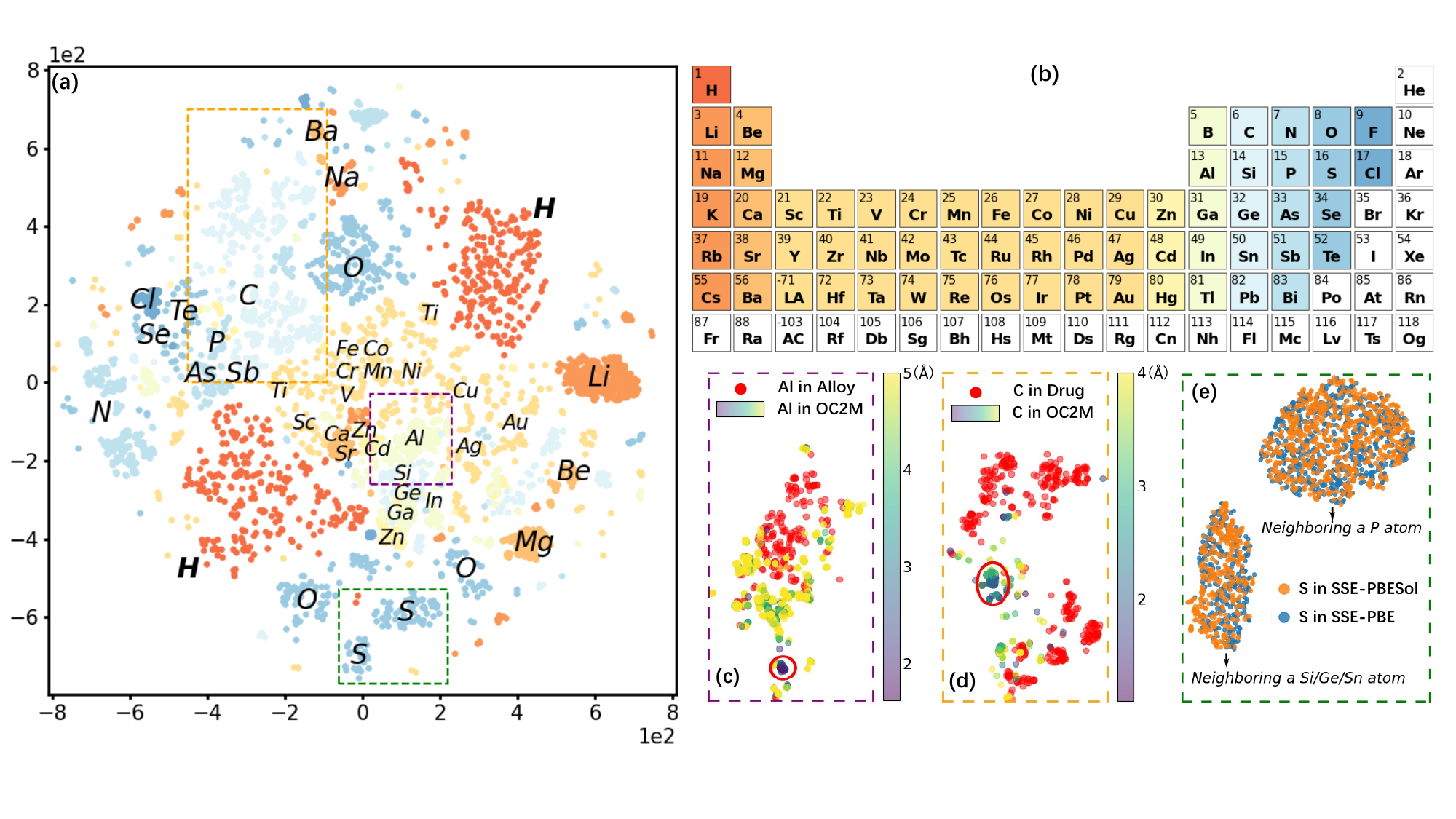}
    \centering
    \caption{t-SNE visualizations of the  DPA-2 single-atom representation of the chemical and configurational space.
    (a-b) Different colors correspond to the different groups in
    the periodic table. From group IA to group VII, red gradually transitions to blue.
    (c) The representations of aluminum in Alloy and OC2M datasets. Red points represent aluminum in Alloy dataset.The gradient colors represent different shortest distances of aluminum in catalyst materials from the adsorbates in the OC2M dataset.
    (d) The representations of carbon in Drug dataset and in adsorbates of the OC2M dataset. Red points represent carbon in Drug dataset. The gradient colors represent different shortest distances of carbon in adsorbates from the catalyst materials in OC2M dataset.
    (e)The representations of sulfur in SSE-PBE and SSE-PBESol datasets.
}
    \label{fig:tsne}
\end{figure*}

We present a visualization of the update of single-atom representations by the final repformer layer using a 2-dimensional t-SNE plot~\cite{Van2008tsne}, as depicted in Fig.\ref{fig:tsne}.
In Fig.\ref{fig:tsne}(a), colors denote distinct groups in the periodic table, as annotated in Fig.\ref{fig:tsne}(b). 
Notably, Fig.\ref{fig:tsne}(a) reveals that representations of identical chemical species tend to form cohesive clusters in the t-SNE latent space.
The distribution of these representations distinctly aligns with known chemistry: 
The elements in groups IA and IIA are clustered at the top right of the t-SNE plot;
The non-metals cluster predominantly at the top left and bottom; 
The transition metals, typically positioned at the middle of the periodic table, are accordingly situated in the central region of the t-SNE figure. 
However, hydrogen (H) presents an exception, exhibiting two clusters: one aligned with metals, primarily in water datasets, and another near non-metals, particularly in molecule datasets such as Drug, ANI-1x, and Transition-1x.

Elements such as Copper (Cu), Silver (Ag), and Gold (Au) in group IB exhibit a tendency to cluster closer to Lithium (Li) than other transition metals due to their shared possession of one s-electron in the outermost electron shell.
Similarly, representations of group IIA elements like Calcium (Ca) and Strontium (Sr) closely associate with those of group IIB elements such as Zinc (Zn) and Cadmium (Cd) owing to their shared possession of two s-electrons in the outermost electron shell. 
Additionally, there's a discernible trend for elements from the same group in the periodic table to cluster together, as evident with Phosphorus (P), Arsenic (As), and Antimony (Sb) from group VII, and Selenium (Se) and Tellurium (Te) from group VIII.

The DPA-2 representation effectively distinguishes between various chemical and configurational environments, as showcased in Fig.\ref{fig:tsne}(c-e).
In Fig.\ref{fig:tsne}(c), representations of Aluminum (Al) atoms from the Alloy and OC2M datasets are depicted. 
The color gradient from purple to yellow indicates the distance of the Al atom from the closest adsorbate in the OC2M dataset, while Al atoms from the Alloy dataset (all-metal environment) are colored red. 
Notably, Al atoms distanced from adsorbates closely resemble those in the Alloy dataset, indicative of similar chemical and configurational environments, whereas those in proximity to adsorbates exhibit discernible differences (see the red-circled blue cluster).
Similarly, Fig.\ref{fig:tsne}(d) illustrates representations of Carbon (C) atoms in the Drug and OC2M datasets. 
Carbon atoms in adsorbates closer to catalyst materials are positioned farther away in latent space from representations in the Drug dataset due to more pronounced differences in their chemical and configurational environments.

Moreover, the DPA-2 representation shows insensitivity to DFT labeling accuracy.
As demonstrated in Fig.~\ref{fig:tsne}(e), representations of sulfur (S) in SSE-PBE (labeled with PBE exchange correlation functional) and SSE-PBESol (labeled with PBE-Sol exchange correlation functional) datasets exhibit mutual overlap. 
The S atoms form two clusters, with one cluster indicating a phosphorus neighboring atom and the other representing a neighboring Si/Ge/Sn atom.

In summary, our analysis reveals that atoms sharing similar chemical and configurational environments are closer in the representation space learned by the DPA-2 model. 
Thus, the DPA-2 representation emerges as a promising candidate for encoding chemical and configurational information in molecular and condensed-phase applications.

\section{Discussion}

In this work, we introduce DPA-2, a Large Atomic Model (LAM), supported by a comprehensive pipeline that includes multi-task pre-training, fine-tuning, knowledge distillation, and practical deployment. The principal findings concerning DPA-2 are as follows:
(1) DPA-2 demonstrates exceptional ability for generalization, primarily due to the multi-task pre-training approach, which utilizes 18 datasets covering 73 chemical elements. These datasets would not typically be merged in a single-task pre-training scenario due to differing labeling methodologies, such as exchange-correlation functionals, energy cutoffs, and k-space grid spacing.
(2) In downstream tasks, the multi-task pre-training approach enables a reduction in data requirements by approximately 1--2 orders of magnitude without sacrificing accuracy.
These results suggest that the DPA-2 model, along with the proposed workflow, stands as a promising framework for molecular and materials simulation.

% perspectives
It is evident that the existing pre-training datasets for the DPA-2 model are insufficient. For example, the datasets currently in use are notably deficient in information on 2-D materials, which significantly limits the model's generalizability to such systems. As a result, the development of LAMs like DPA-2 must be considered a long-term endeavor. This process necessitates the ongoing collection of diverse training data, the incorporation of application-specific test cases, and the establishment of automated workflows for data preprocessing, model training, model evaluation, and version updates.
In recognition of these needs, we underscore the importance of fostering LAMs within an open and collaborative ecosystem. Such an approach would enable the molecular simulation community to both benefit from and contribute to the evolution of LAMs. Reflecting our commitment to this vision, we have launched the OpenLAM Initiative~\footnote{\url{https://deepmodeling.github.io/blog/openlam/}}. Updates on this initiative will be regularly posted on the AIS Square platform~\footnote{\url{https://www.aissquare.com/openlam}}. We cordially invite readers to participate in this project in any capacity they deem fit.
% We emphasize the importance of an open, application-oriented model evaluation system for the molecular simulation community, especially in the era of LAMs. 

%%%%%%%%%%%%%%%%%%%%%%%%%%%%%%%%%%%%%%%%%%%%%%%%%%%%%%%%%%%%

\section{Methods}
\label{sec:method}
% \LZ{need better connected to Fig. 1}

\subsection{Formulation}
\label{sec:formulation}

In this study, we examine a system consisting of $N$ atoms, where the atomic numbers are represented by the list $\mathcal Z = \left(Z_1, \dots, Z_i, \dots, Z_N \right)$, and the atomic coordinates are denoted by the list $\mathcal R = \left({r}_1, \dots, {r}_i, \dots, {r}_N \right)$.
The potential energy surface (PES) of the system is symbolized by $E$, a function dependent on elemental types and coordinates, expressed as $E=E(\mathcal X),\, \mathcal X := ( \mathcal{R}, \mathcal Z)$.
The potential energy surface can be further decomposed into the following equation:
\begin{align}
    E = \sum_i E_i,
\end{align}
where $E_i$ signifies the atomic energy contributions originating from atom~$i$. 
The atomic force exerted on atom $i$, represented as ${F}_i$, is defined as the negative gradient of the total energy with respect to the coordinate: 
\begin{equation}\label{eq:force}
{F}_{i}=-\nabla_{r_i} E.
\end{equation}
For periodic systems, the virial tensor can be obtained as follows:
\begin{align} 
    \Xi_{\alpha\beta}&=-\sum_{\gamma} \frac{\partial E}{\partial h_{\gamma\alpha}} h_{\gamma\beta},
\end{align}
where $\Xi_{\alpha\beta}$ corresponds to the $\alpha\beta$ component of the virial tensor, and $h_{\alpha\beta}$ yields the $\beta$-th component of the $\alpha$-th cell vector.

\subsection{The DPA-2 model}

\subsubsection{The overall architecture of the DPA-2 model}
\label{sec:model_arch}

The DPA-2 is a model that predicts the atomic energy contribution based on the atomic numbers $\mathcal Z$ and the coordinates $\mathcal R$. It consists of two parts,
\begin{align}
    E_i = 
    \mathcal F \Big(  
    \mathcal D_i \big(
    \mathcal R, \mathcal Z
    \big)
    \Big),
\end{align}
where $\mathcal D_i$ represents the descriptor of atom $i$. 
The descriptor must be a smooth mapping from the atomic numbers and coordinates to a hidden representation that remains invariant under translational, rotational, and permutational (only among atoms with the same atomic number) operations.

The fitting network $\mathcal F$ is usually modeled by a standard multiple-layer perceptron (MLP) composed of an energy-biasing layer, 
\begin{align}
    \mathcal F(\mathcal D_i) = 
    e_{\mathrm{bias}} \Big(
    \mathrm{MLP} (\mathcal D_i) 
    \Big).
\end{align}
The energy bias layer ``$e_{\mathrm{bias}}$'' adds a constant bias to the atomic energy contribution according to the atomic number, i.e.,~$e_{\mathrm{bias}}(Z_i) (
    \mathrm{MLP} (\mathcal D_i)) = \mathrm{MLP} (\mathcal D_i) + e_{\mathrm{bias}}(Z_i)$.
Ideally, the energy bias $e_{\mathrm{bias}}$ should be taken as the energy of an atom in a vacuum. 
In practice, the energy bias may be determined by a least-square fitting of the energies in the training data. 
More precisely, suppose we have $M$ data frames, and within the $m$-th frame, we have $c_{mz}$ atoms with atom number $z$, and the DFT labeled energy of the frame is denoted by $E^\ast_{m}$. Then the linear system 
\begin{align}\label{eq:ebias-solve}
    \sum_{z} c_{mz} e_{\mathrm{bias}}(z) = E^\ast_{m}, \quad m=1, \dots, M,
\end{align}
is solved in the least-square sense. 
Here we assume that the number of independent equations in system Eq.~\eqref{eq:ebias-solve} is equal to or smaller than the number of frames $M$.

The DPA-2 descriptor is graphically illustrated in Fig.~\ref{fig:model}, specifically,
\begin{align}
    \mathcal D_i & = 
    \mathrm{concat}(f^0_i, f^2_i),
\end{align}
where $f^0_i$ and $f^2_i$ denote the single-atom representations of atom $i$. 
The requirements for smoothness and symmetry preservation in single-atom representations are identical to those for the descriptor. 
The representation $f^0_i$ is defined as
\begin{align}\label{eq:model-f0}
    f_i^0 = \mathrm{MLP} \left( \mathrm{one\_hot} (Z_i) \right).
\end{align}
The atomic number, $Z_i$, is initially converted into a one-hot representation and subsequently embedded by an MLP.
The output $f_i^0$ is the single-atom hidden representation with dimension $n_1^0$.
The single-atom representation is updated by the $\mathrm{repinit}$ (representation-initializer) layer that encodes the information of local configuration, expressed by the pair-atom representations $g_{ij}^0$ a and  $ h_{ij}^0$, into the single-atom representation.  
\begin{align}
    f^1_i = \mathrm{repinit}(f_i^0, g_{ij}^0, h_{ij}^0).
\end{align}
The feature $f^2_i$ is mapped from single-atom representation and pair-atoms representations $g_{ij}^0$, $ h_{ij}$ by a multiple-layer structure,
\begin{align}\label{eq:model-f2}
    f_i^2 = 
    \underbrace{\mathrm{repformer} \circ \cdots \circ
    \mathrm{repformer}}_{\times 12}
    \Big( 
    \mathrm{linear}(f_i^1),
    \mathrm{linear}(g_{ij}^1),
    h_{ij}^1
    \Big),
\end{align}
where the single- and pair-atom representations are updated by $\mathrm{repformer}$ (representation-transformer) layers. 
The $\mathrm{repformer}$ is designed in a way that the input and output representations share the shape dimension, thus they are stacked 12 times. 
The ``$\circ$''  in Eq.~\eqref{eq:model-f2} thus denotes the layer composition (or mathematically the function composition).
The linear mappings are used to change the dimension of $f_i^1$ and $g_{ij}^1$ to match the shape requirement of $\mathrm{repformer}$.
The pair-atom representations $g_{ij}^0$,  $ h_{ij}^0$, $g_{ij}^1$ and  $ h_{ij}^1$ will be introduced shortly later.
% The $\mathrm{linear}$ layer applies in a representation-wise manner to change the dimension of the representations $f_i^1$ 
% $\mathrm{repinit}$ and $\mathrm{repformer}$ stand for the  and representation-transformer layers, respectively, and are defined in Secs.~\ref{sec:dpa1} and \ref{sec:dpa2}.
It is assumed that the $\mathrm{repinit}$ and $\mathrm{repformer}$ layers only require the information of $i$'s neighboring atoms, i.e., all atoms falling within a sphere centered at atom $i$ with a radius $r_c$.
This radius is commonly referred to as the cut-off radius.
We thus introduce the notation $N_{r_c}(i)$, which represents the set of all neighbors of $i$, i.e., $N_{r_c}(i) = \{ j : j\neq i, \vert  r_j-  r_i\vert < r_c\} $.
The maximum possible number of neighbors for the atoms in the system is denoted by $N_{r_c}^m$, so we have $|N_{r_c}(i)| \leq N_{r_c}^m$, $\forall i$.

To define the pair-atom representations,  $g_{ij}^0$, $ h_{ij}^0$, we consider the local configuration of atom $i$ represented by the augmented environment matrix with shape ${N_{r^0_c}^m \times 4}$, where $r_c^0$ is the cut-off radius used to compute the pair-atom representations.
The $j$-th row of the environment matrix, being a 4-dimensional vector, is defined by
\begin{align}\label{eq:env-mat}
    % \big(\tilde{\mathcal R}_i\big)_{j,\cdot} = 
    \tilde r_{ij} = 
    s(r_{ij}) \times
    \left(1, \frac{x_{ij}}{\vert r_{ij}\vert }, \frac{y_{ij}}{\vert r_{ij}\vert }, \frac{z_{ij}}{\vert r_{ij}\vert }\right),
\end{align}
where $(x_{ij}, y_{ij}, z_{ij})$ are the Cartesian coordinates of the relative position $r_{ij} = r_i -  r_j$.
In most cases, the number of neighbors is smaller than $N_{r_c}^m$, so the environment matrix only has $|N_{r_c}(i)|$ rows defined by Eq.~\eqref{eq:env-mat}, and the remaining positions are filled with zeros.
The switched inverse distance function $s$ in Eq.~\eqref{eq:env-mat} is defined by
\begin{align}\label{eq:s}
    s\left(r_{ij}\right)= \frac{w_{ij}}{\vert r_{ij}\vert },\quad w_{ij} = w(\vert r_{ij}\vert ).
\end{align}

The switch function $w$ takes the value 0 outside the cut-off radius $r_c$, and 1 inside a starting point of switching, denoted by $r_{cs}$. 
In between $r_{cs}$ and $r_c$, the switch function smoothly changes from 1 to 0. 
It is required that $w$ has a continuous second-order derivative on $\mathbb R$. 
One possible implementation of $w$ is provided as 
\begin{align}\label{eq:w}
    w\left(\vert r_{ij}\vert \right)=  
\begin{cases}
1 & \text{if } r_{ij}<r_{c s}, \\ 
u^{3}\left(-6 u^{2}+15 u-10\right)+1 & \text{if } r_{c s} \leq r_{ij}<r_{c},  \\ 
0 & \text{if } r_{c} \leq r_{ij},
\end{cases}
\end{align}
where $u=({\vert r_{ij}\vert -r_{c s}})/({r_{c}-r_{c s}}) $ and $r_{cs} < r_c$ is the starting point of the smooth switch.

The first column of the augmented environment matrix is defined as the rotationally invariant pair-atom representation, while the remaining three columns are denoted by the rotationally equivariant pair-atom representation, i.e.
\begin{align}\label{eq:g0}
    g_{ij}^0 & = s(r_{ij}),\\\label{eq:h}
    h_{ij}^0 & = s(r_{ij}) \times \left(\frac{x_{ij}}{\vert r_{ij}\vert}, \frac{y_{ij}}{\vert r_{ij}\vert}, \frac{z_{ij}}{\vert r_{ij}\vert}\right).
\end{align}
The procedure for calculating pair-atom representations is graphically illustrated in Fig.~\ref{fig:model}(f).
The representations $g_{ij}^1$ and $h_{ij}^1$ are established in precisely the same manner as $g_{ij}^0$ and $h_{ij}^0$, with the only potential variation being the selection of a distinct cut-off radius, denoted as $r_c^1$.

\subsubsection{The repinit layer}\label{sec:dpa1}

The $\mathrm{repinit}$ layer only updates the single-atom $f_i^0$ and pair-atom $g_{ij}^0$ representations, and does not update the equivariant pair-atom representation $h_{ij}$ that is of dimension 3.
The $\mathrm{repinit}$ layer first embeds the concatenated single- and pair-atom representations to update the pair-atom representation
\begin{align}
    g^{rt}_{ij} &= \mathrm{MLP}( \mathrm{concat}(f_i^0, f_j^0, g_{ij}^0)), \quad \forall j \in N_{r_c^0}(i)
    % g^r_{ij} &= \mathrm{MLP}(g_{ij}^0), \\
    % g_{ij}^{rt} &= g^r_{ij} \odot g^t_{ij} + g^r_{ij}.
\end{align}
Then, we concatenate the $g_{ij}^0$ and $h_{ij}$ pair-atom representations to recover the environment matrix and update single-atom representation using a symmetrization operation
\begin{align}
    f_i^1 = \mathrm{linear}(f_i^0) + \mathrm{symm}(g_{ij}^{rt}, \tilde r_{ij}).
\end{align}
The symmetrization operator, first introduced by Ref.~\cite{zhang2018end}, has the general form of $\mathrm{symm}(x_j, y_j)$, where $x_j$ and $y_j$ are neighbor indexed vectors. 
It is assumed that $x_j$ is rotationally invariant, while $y_j$ is not, but the inner product is rotationally invariant.
The symmetrization operator is defined by
\begin{align}\label{eq:symm-2}
    &\mathrm{symm}(x_j, y_j) = 
    \underset{\alpha\gamma}{\mathrm{flatten}}\Big(
    \sum_\beta p_{\alpha\beta} \, p^<_{\gamma\beta}
    \Big),\\\label{eq:symm-0}
    &p_{\alpha\beta}  = \frac 1{N_{r_c^0}^m} 
    \sum_{j\in N_{r_c^0}(i)}
    w_{ij}\, x_{j,\alpha}\, y_{j,\beta}, \\\label{eq:symm-1}
    &p^<_{\alpha\beta} = \underset{\alpha}{\mathrm{split}}(p_{\alpha\beta}).
\end{align}
In Eq.~\eqref{eq:symm-2}, the matrix dimensions $\alpha$ and $\gamma$ are flattened to form a vector.
In Eq.~\eqref{eq:symm-0}, the summation is taken over the index of neighbors $j$, making the matrix $p$ permutationally invariant. 
When an atom comes into the neighborhood of atom $i$, the quantities $x_j$ and $y_j$ generally do not smoothly switch from 0. To prevent the discontinuous jump, the switch $w_{ij}$ is multiplied. 
In Eq.~\eqref{eq:symm-1}, the matrix $p_{\alpha\beta}$ is split along the $\alpha$ dimension, and the first certain number of elements are taken and assigned with notation $p^<$.
It can be proven that the symmetrization operator is invariant with respect to rotational operations and permutational operations over atoms of the same atomic number~\cite{zhang2018end}.

\subsubsection{The repformer layer}\label{sec:dpa2}
The $\mathrm{repformer}$ layer maintains the input and output dimensions of the single- and pair-atom representations, allowing it to be stacked to enhance its representational capabilities.
However, the output of $\mathrm{repinit}$ may not necessarily satisfy the dimension requirements of the $\mathrm{repformer}$ layer. 
To address this issue, the representations are first projected to the desired shape using a linear layer, as follows:
\begin{align}\label{eq:12proj-0}
    &f_i^{2,0} = \mathrm{linear}(f_i^1),\\\label{eq:12proj-1}
    &g_{ij}^{2,0} = \mathrm{linear}(g_{ij}^1),\\
    &h_{ij}^{2,0} = h_{ij}^1.
\end{align}
Subsequently, these representations are updated by the $\mathrm{repformer}$ layers.
The dimensions of the single- and pair-atom representations are denoted by $n^2_1$ and $n^2_2$, respectively.
In the subsequent discussion, the input representations for the $l$-th $\mathrm{repformer}$ layer are denoted by $f_i^{2,l}$ and $g_{ij}^{2,l}$.

In each $\mathrm{repformer}$ layer, the single-atom representation is updated by
\begin{align}\label{eq:dpa2-1atom-0}
    f_i^{2,l+1} = \frac 1{\sqrt 3}
    \bigg(
    f_i^{2,l} + 
    \mathrm{MLP}\big( \tilde  f_i^{2,l} \big) +
    \mathrm{loc\_attn}\big(  f_i^{2,l}    \big)
    \bigg).
\end{align}
The intermediate representation $\tilde f_i^{2,l}$ is defined by
\begin{align}\label{eq:dpa2-1atom-1}
   \tilde f_i^{2,l} = 
   \mathrm{concat}
   \bigg(
   f_i^{2,l}, \:
   \frac 1{N_{r_c^1}^m}\sum_{j\in N_{r_c^1}(i)} 
   w_{ij} g_{ij}^{2,l} \hat f_j^{2,l}, \:
   \mathrm{symm}\big( f_j^{2,l}, h_{ij}^{2,l}\big), \, 
   \mathrm{symm}\big( g_{ij}^{2,l}, h_{ij}^{2,l}\big)
   \bigg),
\end{align}
where $\hat f_j^{2,l}$ is a linearly transformed $f_j^{2,l}$ that has the same dimension as the equivariant pair-atom channel, i.e.~$\hat f_j^{2,l} = \mathrm{linear}(f_j^{2,l})$.
The last term in Eq.~\eqref{eq:dpa2-1atom-0} is the local multi-head self-attention, defined by
\begin{align}
\mathrm{loc\_attn}\big(  f_i^{2,l}    \big)=
\underset{\beta,h\rightarrow n^2_1}{\mathrm{linear}} \Big(\sum_{j\in N_{r_c^1}(i),\alpha} B^{l,\eta}_{ij} f^{l}_{j,\alpha} \hat V^{l,\eta}_{\alpha,\beta} \Big),
\end{align}
with the attention map $B$ given by
\begin{align}
&
\hat q_{i,\gamma}^{l,\eta} = \sum_{\alpha} f^l_{i,\alpha}\, \hat Q^{l,\eta}_{\alpha,\gamma}, \quad
\hat k_{j,\gamma}^{l,\eta} = \sum_{\beta} f^l_{j,\beta}\, \hat K^{l,\eta}_{\beta,\gamma}, \\
\label{eq:dpa2-1atom-attn}
&
B^{l,\eta}_{ij} = \underset{j\in N_{r_c^1}(i)}{\mathrm{softmax}^\ast}
    \Big(
    \frac 1{\sqrt {\hat d}}
    \sum_{\gamma} \hat q_{i,\gamma}^{l,\eta} \hat k_{j,\gamma}^{l,\eta}
    \Big).
\end{align}
Here, $\hat d$ denotes the hidden dimension of the local self-attention, and the $\hat Q$, $\hat K$, and $\hat V$ are trainable matrices. 
The ``$\ast$'' over the softmax operator indicates that the softmax used in Eq.~\eqref{eq:dpa2-1atom-attn} is modified to guarantee the smoothness of the attention map. 
The definition will be introduced in Sec.~\ref{sec:smooth}.

In each layer, the rotationally invariant pair-atom representation is updated by
\begin{align}\label{eq:dpa2-2atom-0}
    \hh^{2,l+1}_{ij} =
    \frac{1}{\sqrt 4} \bigg(
    \hh^{2,l}_{ij} + 
    \mathrm{MLP}(\hh^{2,l}_{ij}) + 
    w_{ij} \underset{n_1^2\rightarrow n_2^2}{\mathrm{linear}} (f_i^{2,l} \odot f_j^{2,l}) +
    \mathrm{gated\_attn} \big(
    g_{ij}^{2,l}, h_{ij}
    \big)\bigg),
\end{align}
where the last term in Eq.~\eqref{eq:dpa2-2atom-0} is the gated multi-head self-attention, which is defined by
\begin{align}\label{eq:dpa2-2atom-1}
    \mathrm{gated\_attn} \big(
    g_{ij}^{2,l}, h_{ij} \big)
    =
    \underset{\beta,h\rightarrow n_2^2}{\mathrm{linear}} \Big(\sum_{k\in N_{r_c^1}(i),\alpha} A^h_{ijk} \hh^{2,l}_{ik,\alpha} V^{l,\eta}_{\alpha,\beta} \Big).
\end{align}
In Eq.~\eqref{eq:dpa2-2atom-1}, the attention map $A$ is given by
\begin{align}
&
q_{ij,\gamma}^{l,\eta} = \sum_\alpha \hh^{2,l}_{ij,\alpha}\, Q^{l,\eta}_{\alpha,\gamma},\quad
k_{ik,\gamma}^{l,\eta} = \sum_\beta  \hh^{2,l}_{ik,\beta}\, K^{l,\eta}_{\beta,\gamma},\\
\label{eq:dpa2-2atom-attn}
&
A^{l,\eta}_{ijk}
    = \underset{k\in N_{r_c^1}(i)}{\mathrm{softmax}}^{\dagger}
    \bigg(
    \Big(
    \frac {1}{\sqrt {d}}
    \sum_{\gamma} q_{ij,\gamma}^{l,\eta} k_{ik,\gamma}^{l,\eta}  
    \Big)
    \:
    \Big(
    \sum_{\delta} \rr_{ij,\delta} \rr_{ik,\delta}
    \Big)
    \bigg),
\end{align}
where $d$ denotes the hidden dimension of the self-attention, the $Q$, $K$, and $V$ are trainable matrices, and $\eta$ is the index of the attention heads.
The gate term $\rr_{ij} \rr_{ik}^T$ is proved to be critical to the generalization ability of the model~\cite{zhang2022dpa}.
As detailed in Sec.~\ref{sec:smooth}, the $\dagger$ over the softmax operator indicates that the softmax used in Eq.~\eqref{eq:dpa2-2atom-attn} is modified to guarantee the smoothness.

We notice that it is fully valid to update the rotationally equivariant representation $\rr_{ij}$ in a similar way, e.g.,
\begin{align}
     \rr^{2,l+1}_{ij} =
    \frac{1}{\sqrt{2}} \bigg(
    \rr^{2,l}_{ij} + 
    \underset{h}{\mathrm{linear}}\Big(
    \sum_{k\in N_{r_c^1}(i)}A^h_{ijk} \rr^{2,l}_{ik}
    \Big)\bigg).
\end{align}
However, we find such an update would not improve the accuracy and often make the training procedure unstable.
Therefore, we choose not to update $\rr_{ij}$ in the current version of the DPA-2 model.

\subsubsection{Smoothness of the softmax operation}\label{sec:smooth}

The standard softmax is defined by
\begin{align}\label{eq:softmax}
    \mathrm{softmax}(x_{ij}) = \frac{e^{x_{ij}}}{\sum_k e^{x_{ik}}},
\end{align}
which introduces discontinuity in the attention maps in Eqs.~\eqref{eq:dpa2-1atom-attn} and \eqref{eq:dpa2-2atom-attn}. 
Simply multiplying a switch to the attention maps does not fix the problem.
Suppose that one atom comes into the cut-off; the denominator of Eq.~\eqref{eq:softmax} changes in a discontinuous way, thus all $\mathrm{softmax}(x_{ij})$ change discontinuously, no matter whether $j$ is the new neighbor or not. 

To fix this issue, we define the $\mathrm{softmax}^\ast$ by
\begin{align}\label{eq:softmax-0}
    \mathrm{softmax}^\ast(x_{ij}) = 
    w_{ij}\, \mathrm{softmax}\big( w_{ij} (x_{ij} + s^\ast)  - s^\ast \big).
\end{align}
Similarly, the $\mathrm{softmax}^\dagger$ is given by
\begin{align}\label{eq:softmax-1}
    \mathrm{softmax}^\dagger(y_{ijk}) = 
    w_{ij} w_{ik}\, \mathrm{softmax}\big( w_{ij} w_{ik} (y_{ijk} + s^\dagger)  - s^\dagger \big).
\end{align}
It is assumed that the shifting constants $s^\ast$ and $s^\dagger$ are chosen a magnitude larger than $x_{ij}$ and $y_{ijk}$, respectively. 
In practice, the magnitude of both $x_{ij}$ and $y_{ijk}$ in Eqs.~\eqref{eq:softmax-0} and \eqref{eq:softmax-1} are of order 1, so we set $s^\ast = s^\dagger = 20$.

\subsection{Single-task training}
\label{sec:st}

Suppose that we have a training dataset $T$ of size $M$, and denote the DFT-labeled energy and force for any configuration $\mathcal{X}_m$, $1\leq m \leq M$, by $E^\ast_m$ and $\{F^\ast_{i,m}\}$, respectively.
The dataset $T$ yields
\begin{align}
    T = \{ (\mathcal{X}_1, E^\ast_1, \{F^\ast_{i,1}\}), \dots, (\mathcal{X}_M, E^\ast_M, \{F^\ast_{i,M}\})   \}.
\end{align}
We denote the trainable parameters of the descriptor by $\theta$, and those of the fitting network by $\xi$. 
When necessary, the parameters are placed as superscripts of the corresponding notation, i.e.,~we have $\mathcal{D}_i^\theta$ and $\mathcal{F}^\xi$ for the descriptor and fitting network, respectively. 
The PES model is thus rewritten as $E = E^{\theta,\xi}(\mathcal{X})$.
The loss function at training step $t$ is written as
\begin{align}
    \mathcal{L}(\theta, \xi, B, t) 
    &=
    \frac{1}{\vert B\vert} \sum_{m\in B}
    \Big(
    \frac {p_e(t)}N \left\vert
    \Delta E_{m}^{\theta,\xi}
    \right \vert^2
    +
    \frac { p_f(t)}{3N} \sum_i \left\vert 
    \Delta F_{i,m}^{\theta,\xi}
    \right \vert^2
    \Big), \\
    \Delta E_{m}^{\theta,\xi}
    &=
    E^{\theta,\xi} (\mathcal{X}_m) - E^\ast_m, \\
    \Delta F_{i,m}^{\theta,\xi}
    &=
    F_{i}^{\theta,\xi} (\mathcal{X}_m) - F^\ast_{i,m},
\end{align}
where $B$, a randomly sampled subset of $\{1, \dots, M\}$, represents the minibatch of the training dataset. 
$p_e(t)$ and $p_f(t)$ are the energy and force prefactors, respectively. 
If the learning rate at step $t$ is denoted by $\gamma(t)$, then the prefactors are defined by
\begin{align}
    p_\xi(t) = p_\xi^{\mathrm{start}} \frac{\gamma(t)}{\gamma(0)} + 
    p_\xi^{\mathrm{limit}} \Big(1-\frac{\gamma(t)}{\gamma(0)}\Big), \quad \xi\in\{e, f\}.
\end{align}
At the beginning of the training, the prefactor $p_\xi$ is set to a hyperparameter $p_\xi^{\mathrm{start}}$, and it linearly decays with respect to the learning rate. 
If the learning rate decays to zero, i.e.,~$\lim_{t\rightarrow\infty}\gamma(t) = 0$, the prefactor converges to the hyperparameter $p_\xi^{\mathrm{limit}}$ at the infinite training step.
We have adopted the Adam stochastic gradient descent method~\cite{kingma2014adam} to minimize the loss function with respect to the model parameters $\theta$ and $\xi$.
Virial errors, which are omitted here, can be added to the loss for training if available.

% For model training or pre-training, we adopted the Adam stochastic gradient descent method~\cite{kingma2014adam} on all the trainable parameters $\boldsymbol{w}$ inside the model to minimize the loss:
% \begin{equation}
% \label{loss:single}
%     \mathcal{L}_{\boldsymbol{w}}(E^{\boldsymbol{w}}, \mathcal{F}^{\boldsymbol{w}}) = \frac{1}{\vert\mathcal{B}\vert}\sum_{t \in \mathcal{B}}
%     \left( 
%     p_{\epsilon}\left \vert E_{t}-E_{t}^{\boldsymbol{w}}\right\vert^{2}
%     + p_{f}\left\vert\mathcal{F}_{t}-\mathcal{F}_{t}^{\boldsymbol{w}}\right\vert^{2} 
%     \right).
% \end{equation}
% Here $\mathcal{B}$ represents a minibatch, $\vert\mathcal{B}\vert$ is the batch size, $t$ denotes the training sample. 
% $E^{\boldsymbol{w}}_{t}, \mathcal{F}^{\boldsymbol{w}}_{t}$ denote the model outputs on $t$ and $E_t, \mathcal{F}_t$ are the corresponding DFT results. We also adopted a scheduler to tune the prefactors $p_{\epsilon}$ and $p_{f}$ during the training process to make a better balance between energy and force labels. 
% Virial errors, which are omitted here, can be added to the loss for training if available.

\subsection{Multi-task training protocol}
\label{sec:mt}

% \begin{figure*}
%     \includegraphics[width=1.0\textwidth]{figs/multitask.pdf}
%     \centering
%     \caption{Multi-task training procedure in DPA-2.
%     }
%     \label{fig:multi-task}
% \end{figure*}

%In the Sec.~\ref{sec:st}, the DPA-2 model is trained on data calculated with consistent DFT parameters, which was the focus of most previous studies.
For various datasets labeled with different DFT calculation parameters, it is infeasible to merge them directly into a single training set for model training. However, these DFT datasets should inherently share a significant amount of commonality, and we expect they can mutually promote each other's training, thus benefiting the overall model capacity.

In this work, to fully utilize various sources of DFT calculated data, we propose a novel \emph{multi-task} training strategy using a unified model framework for simultaneous training on data calculated with different DFT parameters, as illustrated in Fig.~\ref{fig:pipeline}(a).
We first group all the training data into $K$ training datasets, denoted as $\mathcal T = \{T_1, \dots, T_K\}$, where each dataset contains configurations labeled with identical DFT parameters.
The configurations and labels in the $k$-th training dataset are represented by:
\begin{align}
    T_k = \{ (\mathcal X_{k1}, E^\ast_{k1}, \{F^\ast_{i,k1}\}), \dots, (\mathcal X_{kM}, E^\ast_{kM}, \{F^\ast_{i,{kM}}\})   \}.
\end{align}

We establish a DPA-2 model with the unified descriptor and $K$ fitting networks, and the $k$-th model is given by:
\begin{align}
    E = E^{\theta,\xi_k}(\mathcal X),
\end{align}
where $\xi_k$ represents the network parameters of the $k$-th fitting network.
The $k$-th fitting network is trained by the $k$-th training dataset, while the unified descriptor (with parameters $\theta$) is \emph{simultaneously} trained by all datasets, and the loss function is given by
\begin{align}
    \mathcal L(\theta, \{\xi_k\}, S, \{B\}, t) 
    &=
    \frac{1}{\vert S\vert} 
    \sum_{k\in S}
    \frac{1}{\vert B_k\vert} 
    \sum_{m\in B_k}
    \Big(
    \frac {p_e(t)}{N_m} \left\vert
    \Delta E_{km}^{\theta,\xi_k}
    \right \vert^2
    +
    \frac { p_f(t)}{3N_m} \sum_i \left\vert 
    \Delta F_{i,km}^{\theta,\xi_k}
    \right \vert^2
    \Big), \\
    \Delta E_{km}^{\theta,\xi_k}
    &=
    E^{\theta,\xi_k} (\mathcal X_{km}) - E^\ast_{km}, \\
    \Delta F_{i,km}^{\theta,\xi_k}
    &=
    F_{i}^{\theta,\xi_k} (\mathcal X_{km}) - F^\ast_{i,km}.
\end{align}
At each training step, a subset of the training datasets is sampled from $\mathcal T$, and the indices of the sampled datasets are denoted by $S$.
$B_k$ represents the minibatch of the training dataset $T_k$. 
It should be noted that there is a significant degree of freedom in designing the sampling strategy for $S$. 
Sampling can be conducted with a uniform probability or with a bias towards certain systems.
Furthermore, sampling may be performed with or without replacement.
In our implementation, larger and more complex datasets are assigned a higher probability, and sampling with replacement is employed.

% Regarding the training process, we can control the drawing probabilities $\bm p$ by setting weights of each dataset. In each training step, we can only update the descriptor and the fitting net corresponding to the currently chosen dataset. In a multi-GPU parallel scenario, we can even train different datasets with their losses on different GPUs in a single training step, which is much more efficient. Expecting the unified framework to obtain a more general representation in the descriptor, we hope that various datasets can help each other, and under the premise of controlling the training steps of each dataset to be the same as that of the single-task model, we aim to achieve higher accuracy.

\subsection{Pre-training and fine-tuning}

By utilizing multi-task training on all available training datasets, the configurational and elemental knowledge shared among the datasets is expected to be encoded in the descriptor $\mathcal D^{\theta_p}$, with $\theta_p$ denoting the converged model parameters. The fitting networks are expected to encode system-specific knowledge. 
The multi-task training scheme provides the possibility of training with a large number of training datasets (most likely labeled with distinct DFT parameters). 
Therefore, when trained with a sufficiently large dataset that covers a wide range of configurations and elements for future applications, it is expected that much less training data would be needed to train a new system with the help of the encoded knowledge. 
The multi-task \emph{pre-trained} model can be used to improve the accuracy and data efficiency in \emph{downstream tasks}. It is worth noting that the downstream task can be either constructing a PES, or a property prediction task, and in this work, we only discuss the PES as a downstream task. The procedure of training a model for downstream tasks from a pre-trained model is called \emph{fine-tuning}.

Given a downstream task training dataset, we may initialize the descriptor of our downstream task model with $\theta_p$ to boost the performance compared to a random initialization of the descriptor parameters. Furthermore, if the downstream dataset shares similar configurational and elemental information with any of the fitting networks, then the fitting network of the model could also be initialized with the pre-trained fitting network. The energy bias of the downstream task is determined by the downstream training dataset, rather than by those used in the pre-training stage.
% \WH{Here the replacement of energy bias should be explained. before doing that the energy bias should be explained in the model structure sections.}

% Upon obtaining a pre-trained model, we hope that it can benefit from the pre-training in various downstream tasks and exhibit better accuracy performance compared to training from scratch. In the case of single-task pre-training, we directly select the parameters $\boldsymbol{w}$ obtained from pre-training. In the multi-task scenario, we can either choose the parameter branch $\boldsymbol{w}({\bm x^{*}})$ corresponding to the pre-trained dataset ${\bm x^{*}}$ that is most similar to the downstream dataset, or choose to retain only $\boldsymbol{w_d}$ and randomly initialize the $\boldsymbol{w_f}$ part. Following either of these approaches, we then need to modify the energy bias of the last layer in the fitting net to match the new statistical results. Subsequently, we can choose to fix some parameters in the pre-trained model and train the remaining ones.

\subsection{Model distillation}
\label{sec:distill}

The fine-tuned model possesses a large number of parameters, which might result in low efficiency when directly applied to production scenarios, such as MD simulations. To mitigate this issue, we can distill the model into a more compact version that maintains accuracy on downstream tasks while concurrently achieving speed enhancements and enabling large-scale simulations. The distillation process, illustrated in Fig.~\ref{fig:pipeline}(c), consists of an iterative concurrent learning loop. The model prior to distillation, denoted as the teacher model, is used for data labeling, whereas a student model featuring a simpler model structure (e.g., DPA-1 without any attention layer, which can be further compressed~\cite{lu2022dp} to significantly enhance performance) is trained on the labeled data. Subsequently, the teacher model is utilized for MD exploration, adopting simulation settings similar to those of downstream tasks, ensuring that the elemental and configurational spaces explored during distillation and downstream tasks exhibit overlap. Configurations are sampled from the simulated MD trajectories, and the inference deviations between the teacher and student models on those samples are assessed. Samples with model deviation exceeding a predetermined threshold are added to the training dataset for the next iteration. This procedure is repeated until the student model's accuracy satisfies our criteria or no longer changes.

\section{Data and Code Availability}
The datasets and models used in this study, as detailed in Sec.~\ref{sm:sec:data} of the Supplementary Materials, are all available on AIS Square (\url{https://www.aissquare.com}). 
The codes, datasets and input scripts are all available on zenodo (\url{https://doi.org/10.5281/zenodo.10428497}).
Finally, to test the models, users are welcome to consider going through this Bohrium Notebook (\url{https://nb.bohrium.dp.tech/detail/18475433825}), and explore the DP Combo web server (\url{https://app.bohrium.dp.tech/dp-combo}).

\section{Acknowledgements}
We gratefully acknowledge the support received for this work. 
The work of Han Wang is supported by the National Key R\&D Program of China (Grant No.~2022YFA1004300) and the National Natural Science Foundation of China (Grant No.~12122103). 
The work of Weinan E is supported by the National Key Research and Development Project of China (Grant No.~2022YFA1004302) and the National Natural Science Foundation of China (Grants No.~92270001 and No.~12288101).
The work of Jinzhe Zeng and Darrin M. York is supported by the National Institutes of Health (Grant No.~GM107485 to D.M.Y.) and the National Science Foundation (Grant No.~2209718 to D.M.Y.). 
The work of Shi Liu is supported by the Natural Science Foundation of Zhejiang Province (Grant No.~2022XHSJJ006). 
The work of Tong Zhu is supported by the National Natural Science Foundation of China (Grants No.~22222303 and No.~22173032).
The work of Zhicheng Zhong is supported by the National Key R\&D Program of China (Grants No.~2021YFA0718900 and No.~2022YFA1403000).
The work of Jian Lv is supported by the National Natural Science Foundation of China (Grants No.~12034009 and No.~91961204).
The work of Jun Cheng is supported by the National Science Fund for Distinguished Young Scholars (Grant No.~22225302), Laboratory of AI for Electrochemistry (AI4EC), and IKKEM (Grants No.~RD2023100101 and No.~RD2022070501). 
The work of Mohan Chen is supported
by the National Natural Science Foundation of China (Grants No.~12122401, No.~12074007, and No.~12135002).
The work of Yifan Li is supported by the “Chemistry in Solution and at Interfaces” (CSI) Center funded by the United States Department of Energy under Award No.~DE-SC0019394.
Lastly, the computational resource was supported by the Bohrium Cloud Platform at DP Technology and Tan Kah Kee Supercomputing Center (IKKEM).

\newpage

\renewcommand{\thesection}{S\arabic{section}}    %%%% but here
\setcounter{section}{0}
\renewcommand{\thefigure}{S\arabic{figure}}
\setcounter{figure}{0}
\renewcommand{\thetable}{S\arabic{table}}
\setcounter{table}{0}

\section*{Supplementary Materials}

\section{Datasets}
\label{sm:sec:data}
The datasets denoted by an asterisk (\textbf{*}) are being published for the first time in this paper. All of 25 datasets, along with the training and test splits for both pre-training and downstream tasks, can be accessed through the AIS-Square platform at \url{https://www.aissquare.com}.

\begin{itemize}
\item[\textbf{Alloy*\href{https://www.aissquare.com/datasets/detail?pageType=datasets&name=Alloy_DPA_v1_0&id=147}{$^{\mathrm{url}}$}}]
This dataset is generated using the DP-GEN~\cite{zhang2020dp} scheme and comprises structure-energy-force-virial data for 53 typical metallic elements (Li, Be, Na, Mg, Al, Si, K, Ca, Sc, Ti, V, Cr, Mn, Fe, Co, Ni, Cu, Zn, Ga, Ge, Sr, Y, Zr, Nb, Mo, Ru, Rh, Pd, Ag, Cd, In, Sn, La, Hf, Ta, W, Re, Os, Ir, Pt, Au, Pb, Ce, Pr, Nd, Sm, Gd, Tb, Dy, Ho, Er, Tm, Lu) 
%\WH{How many? list all if possible}(e.g., rare earth elements). 
%These elements include matrix materials \WH{How many? better to list all} (e.g., Al, Mg, Ti, Fe) and additive components  for alloys.
The dataset encompasses a diverse array of crystal configurations, featuring FCC, BCC, and HCP structures, as well as intermetallic compounds and amorphous structures with stochastic vacancies. The dataset contains three categories, random substitutional solid solutions, elementary substances, and intermetallic compounds.
%Except for intermetallic compounds, other configurations are substitutional solid solutions, with sites randomly occupied by the metallic elements.
%is consist of The dataset spans a broad composition space, with a focus on  and the exclusion of interstitial elements (e.g., C, B, O, H). 
% We can bifurcate the data into two distinct sets: one consisting of relaxed and systematically deformed data, and the other encompassing randomly distorted data generated by the DP-GEN software.
% The latter, containing a wealth of information during MD simulations, is selected for \textbf{pre-training}. 
% The former is subsequently partitioned into \textbf{downstream} data for fine-tuning purposes.
All density functional theory (DFT) calculations were conducted using the ABACUS package~\cite{chen2010systematically, li2016large}. The exchange-correlation functional was described by the generalized gradient approximation (GGA) in the Perdew-Burke-Ernzerhof (PBE) form. Norm-conserving pseudopotentials were adopted. The cutoff energy of the plane wave basis was set to be 100 Rydberg, and the Monkhorst-Pack~\cite{perdew1996generalized} $k$-point mesh was chosen with a reciprocal space resolution of 0.05 Bohr$^{-1}$.
The self-consistent field iteration stops when the difference in total electron density of consecutive iterations is less than $10^{-6}$.

\item[\textbf{Cathode*\href{https://www.aissquare.com/datasets/detail?pageType=datasets&name=Cathode\%28Anode\%29_DPA_v1_0&id=130}{$^{\mathrm{url}}$}}]
This dataset explores O3-type layered oxide cathodes employed in lithium-ion and sodium-ion batteries.
It has been generated utilizing the DP-GEN scheme.
Specifically, the systems analyzed include \ce{Li_xTMO2} and \ce{Na_xTMO2}, where \ce{TM} represents transition metal elements including Ni, Mn, Fe, Co, and Cr. 
The configuration space is explored by NPT MD simulations in a wide range of temperatures and pressures, varying from 50.0 K to 1250.0 K and 0 bar to 3000 bar, respectively.
The DFT calculations for this dataset were conducted using the VASP~\cite{kresse1996efficiency,kresse1996efficient} software, incorporating the PBE-GGA functional. 
The dataset comprises supercells containing twelve formula units for various systems, including \ce{LiTMO2}, \ce{NaTMO2}, \ce{Li_{0.5}TMO2}, \ce{Na_{0.5}TMO2}, and \ce{TMO2}.
The dataset is split into two non-overlapping subset, \textbf{Cathode-P} and \textbf{Cathode-D}, denoting the pre-training and downstream dataset, respectively. 
The  Cathode-P dataset is composed by  \ce{Li_xTMO2} with TM $\in$ \{Mn, Fe, Co, Cr\}, \ce{Na_xMnO2}, and all \ce{TMO2} with TM $\in$ \{Ni, Mn, Fe, Co, Cr\}, 
while the Cathode-D dataset includes the \ce{Li_xNiO2} and \ce{Na_xTMO2} with TM $\in$ \{Ni, Fe, Co, Cr\}.
% Systems chosen for \textbf{pre-training} include \ce{Li_xTMO2} with TM $\in$ \{Mn, Fe, Co, Cr\}, \ce{Na_xTMO2} with TM = Mn, and all \ce{TMO2}. Meanwhile, other systems of \ce{Li_xTMO2} with TM = Ni and \ce{Na_xTMO2} with TM $\in$ \{Ni, Fe, Co, Cr\} are selected for \textbf{downstream} tasks.

\item[\textbf{Cluster*\href{https://www.aissquare.com/datasets/detail?pageType=datasets&name=Cluster_DPA_v1_0&id=131}{$^{\mathrm{url}}$}}]
This dataset is composed of metal nano-clusters. 
The dataset is decomposed in a non-overlapping way into two subsets, \textbf{Cluster-P} and \textbf{Cluster-D}, which are used for pre-training and downstream tasks, respectively. 
The Cluster-P dataset encompasses 9 types of clusters that are composed of one element, namely Au, Ag, Al, Cu, Ni, Pt, Pd, Si, and Ru, and 15 types of clusters composed of a combination of 2 elements, namely AgCu, AgNi, AgPd, AgPt, AuAg, AuCu, AuNi, AuPd, AuPt, CuNi, CuPd, CuPt, NiPd, PtNi, and PtPd. 
The Cluster-D dataset includes 7 types of clusters composed of ternary combination of metal elements, i.e.~AgCuPt,  AuAgCu,  AuAgPd,  AuAgPt,  AuCuPd,  AuCuPt, and  PtPdNi. 
% containing 31 metal clusters, such as Au, Ag, Al, Cu, Ni, Pt, Pd, Si, and Ru, along with their unitary and various binary or ternary combinations. 
The configurations of the clusters in the Cluster-P and Cluster-D datasets are explored using the DPGEN scheme. 
The DFT calculations are performed using CP2K~\cite{kuhne2020cp2k} with PBE exchange-correlation functional and Grimme D3 dispersion correction~\cite{grimme2010consistent}. 
% All unitary and binary systems have been selected as \textbf{pre-training} data, while ternary systems for \textbf{downstream} tasks.

\item[\textbf{Drug*\href{https://www.aissquare.com/datasets/detail?pageType=datasets&name=Drug\%28drug-like-molecule\%29_DPA_v1_0&id=143}{$^{\mathrm{url}}$}}]
This dataset is generated using the DPGEN approach and encompasses an extensive collection of over 1.4 million structures, comprising 8 elements  H, C, N, O, F, Cl, S, and P, with the inclusion of up to 70 heavy atoms.
The foundation for the initial training data was established by optimizing small molecules procured from the ChEMBL~\cite{jupp2014ebi, bento2014chembl, davies2014mychembl} database with the aid of Gaussian software~\cite{frisch2016gaussian}. 
To expand the dataset, high-temperature simulations were employed, and the data pool was further augmented by optimizing larger molecules from the ChEMBL database, followed by conducting supplementary simulations.
In addition, unoptimized structures were randomly selected and subjected to simulations, resulting in the enlargement of the training set to encompass over 1 million conformations.
To ensure comprehensive torsion coverage, structures originating from the ChEMBL torsion scans dataset were optimized and simulated, while enhanced sampling MD simulations performed with molecules contributed additional structures to the dataset.
The whole dataset is employed for \textbf{pre-training}.

\item[\textbf{FerroEle~\cite{wu2023universal}\href{https://www.aissquare.com/datasets/detail?pageType=datasets&name=FerroEle_DPA_v1_0&id=141}{$^{\mathrm{url}}$}}] This dataset comprises 26 ABO3-type perovskite oxides, which span an extensive composition space containing elements such as Pb, Sr, Ba, Ca, Bi, K, Na, and their various combinations for the A-site, in addition to Ti, Nb, Zr, Mg, Zn, In, Hf, and their respective combinations for the B-site. 
The configurations of the materials were generated utilizing the DPGEN method and subsequently employed for the training of a universal interatomic potential for perovskite oxides, referred to as UniPero. 
All DFT calculations were executed with the aid of the ABACUS software~\cite{chen2010systematically, li2016large}, utilizing the PBEsol functional within the GGA framework and ONCV multi-projector pseudopotentials. This dataset is divided into four distinct segments according to data complexity, as detailed in~\cite{wu2023universal}.
The dataset is divided without overlap into \textbf{FerroEle-P} and \textbf{FerroEle-D} subsets, used for pre-training and downstream tasks, respectively. 
The FerroEle-P dataset is composed by \ce{BaTiO_3}, \ce{CaTiO_3}, \ce{PbTiO_3}, \ce{SrTiO_3}, \ce{NaNbO_3}, \ce{Pb(Zn_{1/3}Nb_{2/3})O_3}, \ce{Pb(Zr_{1-x}Ti_{x})O_3}, \ce{Pb(Mg_{1/3}Nb_{2/3})O_3}, \ce{Pb(In_{1/2}Nb_{1/2})O_3}, \ce{Bi_{1/2}Na_{1/2}TiO_3}, \ce{K_{1/2}Na_{1/2}NbO_3}, \ce{Ba_{x}Ca_{1-x}TiO_3}, \ce{Ba_{x}Sr_{1-x}TiO_3}, \ce{Ca_{x}Sr_{1-x}TiO_3}, \ce{Ba_{x}Pb_{1-x}TiO_3}, \ce{Ca_{x}Pb_{1-x}TiO_3}, and \ce{Pb_{x}Sr_{1-x}TiO_3}.
The FerroEle-D dataset contains perovskite oxide binary solid solutions, 
\ce{Bi_{1/2}Na_{1/2}TiO_3}-\ce{BaTiO_3}, 
\ce{NaNbO_3}-\ce{-BaTiO_3},
\ce{Pb(Mg1/3Nb2/3)O3}-\ce{BaTiO3},
\ce{Pb(Mg1/3Nb2/3)O3}-\ce{PbTiO3},
\ce{Pb(Zn1/3Nb2/3)O3}-\ce{PbTiO3},
\ce{Ba(Zr_{0.2}Ti_{0.8})O_3}-\ce{Ba_{0.7}Ca_{0.3}TiO_3}, \ce{Ba(Hf_{0.2}Ti_{0.8})O_3}-\ce{Ba_{0.7}Ca_{0.3}TiO_3},
and ternary solution 
\ce{Pb(In1/2 Nb1/2 )O3}–\ce{Pb(Mg1/3 Nb2/3 )O3}–\ce{PbTiO3},
\ce{Pb(Mg1/3 Nb2/3 )O3}-\ce{Pb(Zn1/3 Nb2/3 )O3}-\ce{PbTiO3}.
% \WH{ \@wujing plz check the dataset!}
% In general terms, higher levels encompass a greater number of elements in one system. The two levels of lesser complexity are selected for \textbf{pre-training}, while the two more challenging levels for \textbf{downstream}.

\item[\textbf{OC2M~\cite{chanussot2021open}\href{https://www.aissquare.com/datasets/detail?pageType=datasets&name=Open_Catalyst_2020\%28OC20_Dataset\%29&id=50}{$^{\mathrm{url}}$}}] This dataset constitutes a subset derived from the Open Catalyst Project's comprehensive OC20 dataset, which is inclusive of approximately 2 million DFT data samples. This particular dataset comprises 56 distinct elements, with the samples depicting DFT relaxations associated with molecular adsorptions on various surfaces, spanning an extensive structure and chemical space. The principal focus of these samples is directed toward 82 adsorbates that hold significance in the context of renewable energy production and environmental applications. The entire dataset is utilized for \textbf{pre-training}.

\item[\textbf{SSE-PBE~\cite{huang2021deep}\href{https://www.aissquare.com/datasets/detail?pageType=datasets&name=SSE-PBE_DPA_v1_0&id=146}{$^{\mathrm{url}}$}}]
This dataset comprises solid-state electrolytes generated through the DP-GEN method.
It is composed of three distinct chemical formulas, namely \ce{Li10GeP2S12}, \ce{Li10SiP2S12}, and \ce{Li10SnP2S12}. 
All DFT calculations were conducted employing the VASP software~\cite{kresse1996efficiency,kresse1996efficient}, with the application of PBE exchange-correlation functional.
The \ce{Li10GeP2S12}, \ce{Li10SiP2S12} configurations form the \textbf{SSE-PBE-P} dataset used for pre-training, while the \ce{Li10SnP2S12} configurations provide the \textbf{SSE-PBE-D} dataset for downstream tasks.
% The complete dataset serves for the \textbf{pre-training}.

\item[\textbf{SSE-PBESol~\cite{huang2021deep}\href{https://www.aissquare.com/datasets/detail?pageType=datasets&name=SSE-PBESol_DPA_v1.0&id=134}{$^{\mathrm{url}}$}}]
This dataset comprises solid-state electrolyte configurations labeled DFT calculations using the PBESol exchange-correlation functional.
Three formulas, namely \ce{Li10GeP2S12}, \ce{Li10SiP2S12}, and \ce{Li10SnP2S12}, are considered, and the configurations were with generated through a DP-GEN procedure that is independent of that used in SSE-PBE. 
The complete dataset serves for the \textbf{downstream} tasks.

\item[\textbf{SemiCond~\cite{liu2023machinelearningbased}\href{https://www.aissquare.com/datasets/detail?pageType=datasets&name=SemiCond_DPA_v1_0&id=142}{$^{\mathrm{url}}$}}] 
This dataset encompasses 20 semiconductors spanning from group IIB to VIA, namely Si, Ge, SiC, BAs, BN, AlN, AlP, AlAs, InP, InAs, InSb,
GaN, GaP, GaAs, CdTe, InTe-In2Te3, CdSe-CdSe2, InSe-In2Se3, ZnS, CdS-CdS2.
The configurations are explored by the DPGEN scheme in a temperature range of approximately 50.0 K to $\sim$4000 K and a pressure range of circa 1 bar to 50000 bar.
% comprising 2 unitary and 18 binary alloy systems.
% such as \ce{Si}, \ce{Ge}, \ce{SiC}, \ce{In2Se3}, among others.
DFT calculations, employed during the DP-GEN process, are computed utilizing the ABACUS software package~\cite{chen2010systematically, li2016large}. 
The energy cutoff of the DFT calculations was set to 100~Ry ($\sim 1361$~eV) and the mesh grid for K-space sampling was 0.08~Bohr$^{-1}$ ($\sim 0.15$~\AA$^{-1}$).
The dataset is divided into two non-overlapping subsets, \textbf{SemiCond-P} and \textbf{SemiCond-D}, for pre-training and downstream tasks. 
The Ge and Si are placed in the SemiCond-P dataset. 
The 11 of the two-element semiconductors, namely AlAs, AlP,  BN,  CdSe-CdSe2,  CdS-CdS2, CdTe, GaAs, GaN, GaP, SiC, and ZnS are assigned to the SemiCond-P dataset, while the reset semiconductors, AlN, BAs, InAs, InP, InSb, InSe and InTe are assigned to the SemiCond-D dataset.
% For \textbf{pre-training}, we select all the 2 unary systems and randomly choose 11 binary systems, while reserving the remaining 7 binary systems for \textbf{downstream} testing. We intentionally exclude \ce{In2Se3} from pre-training due to its presence in other downstream datasets.

\item[\textbf{\ce{H2O}-PD~\cite{zhang2021phase}\href{https://www.aissquare.com/datasets/detail?pageType=datasets&name=H2O-PD_DPA_v1_0&id=137}{$^{\mathrm{url}}$}}]
The water/ice dataset is used to train a DP model for the calculation of the phase diagram of water in the thermodynamic range of 0 to 2400 K and  0 to 50 GPa. 
The dataset was labeled by the VASP software~\cite{kresse1996efficiency,kresse1996efficient} with the SCAN exchange-correlation functional. 
The energy cutoff was set to 1500~eV and the spacing of the K-space lattice was $0.5$~\AA$^{-1}$.
This dataset is used for pre-training.
% This dataset encompasses DFT samples that span the phase diagram of water, with temperatures ranging from 0 to 2400 K and pressures from 0 to 50 GPa, specifically selected for the purpose of \textbf{pre-training}.

\item[\textbf{Ag$\cup$Au-PBE~\cite{wang2021generalizable}\href{https://www.aissquare.com/datasets/detail?pageType=datasets&name=AgAu-PBE\%28unitary\%29_DPA_v1_0&id=152}{$^{\mathrm{url}}$}}] 
This dataset contains Ag and Au configurations that were generated by the DP-GEN scheme.
% DFT samples through the DP-GEN method.
DFT calculations were conducted employing VASP software~\cite{kresse1996efficiency,kresse1996efficient}, in conjunction with PBE functional. 
The entire dataset is used for \textbf{pre-training}.

\item[\textbf{AgAu-PBED3~\cite{wang2021generalizable}\href{https://www.aissquare.com/datasets/detail?pageType=datasets&name=AgAu-PBED3_DPA_v1.0&id=154}{$^{\mathrm{url}}$}}] 
This dataset contains Ag, Au, and AgAu alloy configurations that were generated by the DP-GEN scheme.
The labels were generated by VASP~\cite{kresse1996efficiency,kresse1996efficient} using the PBE exchange-correlation function in conjugate with the D3 dispersion correction~\cite{grimme2010consistent}.
% DFT samples through the DP-GEN method.
% DFT calculations were conducted employing VASP software, in conjunction with PBE functional. 
The entire dataset is used for \textbf{downstream} tasks.

\item[\textbf{AlMgCu~\cite{jiang2021accurate}\href{https://www.aissquare.com/datasets/detail?pageType=datasets&name=AlMgCu_DPA_v1_0&id=139}{$^{\mathrm{url}}$}}] 
This dataset contains unitary, binary, and ternary alloys of Al, Cu, and Mg, i.e.~\ce{Al_{x}Cu_{y}Mg_{z}} with a concentration range of $0 \leq x,y,z \leq 1,x+y+z=1$. 
% This dataset, generated using DP-GEN, explores 2.73 billion alloy configurations of \ce{Al_{x}Cu_{y}Mg_{z}} systems within a concentration range of $0 \leq x,y,z \leq 1,x+y+z=1$, 
The configurations are explored by the DP-GEN scheme in a temperature range of 50.0 K to 2579.8 K and a pressure range of 1 to 50,000~bar (5~GPa).
% The dataset covers a pressure range of 1 to 50,000~bar (5~GPa), consisting of approximately 100k labeled configurations.
% \WH{write the settings on the label generation}
Energy, force, and virial labels are obtained by DFT calculations adopting the PBE fucntional using VASP~\cite{kresse1996efficiency,kresse1996efficient}. The energy cut-off for PAW basis sets is 650 eV. K-points are sampled by Monkhorst-Pack mesh with a grid spacing of 0.1~\AA$^{-1}$. Order 1 Methfessel-Paxton smearing is used with $\sigma$ = 0.22~eV. SCF convergence criterion for DFT calculation is $1 \times 10^{-6}$~eV.
The dataset is divided into \textbf{Al$\cup$Mg$\cup$Cu} for pre-training and \textbf{AlMgCu-D} for downstream in a mutually exclusive way. 
The Al$\cup$Mg$\cup$Cu dataset contains all the unitary systems, i.e.~pure Al, Mg, and Cu, while the AlMgCu-D contains binary and ternary alloy configurations. 
% For \textbf{pre-training}, solely unitary systems are employed, while both binary and ternary systems are incorporated into the \textbf{downstream} dataset to pose a greater challenge.

\item[\textbf{ANI-1x~\cite{smith2018less}\href{https://www.aissquare.com/datasets/detail?pageType=datasets&name=ANI1x_DPA_v1_0&id=191}{$^{\mathrm{url}}$}}] 
This dataset is generated through iterative active learning to efficiently sample chemical space relevant for machine-learned potentials. It contains over 5 million conformations of organic molecules with up to 13 heavy atoms.
Initial data came from small GDB-11~\cite{fink2005virtual, fink2007virtual} molecules. More complex chemical space was then explored by conformational sampling of progressively larger molecules from databases like ChEMBL and GDB-13~\cite{blum2009970}. Techniques included diverse normal mode sampling, trajectory sampling, and dimer sampling to capture molecular diversity. By combining automated active learning of molecules and conformations with rigorous sampling methods, ANI-1x provides comprehensive coverage of organic chemical space. The entire dataset is used for \textbf{downstream} tasks.

\item[\textbf{Transition-1x~\cite{schreiner2022transition1x}\href{https://www.aissquare.com/datasets/detail?pageType=datasets&name=Transition1x_DPA_v1_0&id=190}{$^{\mathrm{url}}$}}] 
The dataset comprises over 9.6 million conformations of organic small molecules, spanning more than 10,000 distinct organic chemical reactions. Each reaction involves up to seven heavy atoms, including carbon, nitrogen, and oxygen. Originating from the GDB-7~\cite{ruddigkeit2012enumeration} database, the dataset selects structures that serve as reactants; potential products are then generated via the Growing String Method. Reaction trajectories for these reactant-product pairs are computed employing the Nudged Elastic Band (NEB) method. By selectively sampling structures produced during the NEB procedure and discarding non-physical conformations, the dataset effectively encapsulates the chemical space pertinent to reaction pathways. The entire dataset is used for \textbf{downstream} tasks.

\item[\textbf{Other datasets}] To provide a clearer understanding, detailed descriptions of the datasets can be found in the respective links attached to them.
Additional datasets Cu~\cite{zhang2020dp}\href{https://www.aissquare.com/datasets/detail?pageType=datasets&name=Cu_DPA_v1_0&id=132}{$^{\mathrm{url}}$}, Sn~\cite{chen2023modeling}\href{https://www.aissquare.com/datasets/detail?pageType=datasets&name=Sn_DPA_v1_0&id=129}{$^{\mathrm{url}}$}, Ti~\cite{wen2021specialising}\href{https://www.aissquare.com/datasets/detail?pageType=datasets&name=Ti_DPA_v1_0&id=133}{$^{\mathrm{url}}$}, 
V~\cite{wang2022classical}\href{https://www.aissquare.com/datasets/detail?pageType=datasets&name=V_DPA_v1_0&id=135}{$^{\mathrm{url}}$}, W~\cite{wang2022tungsten}\href{https://www.aissquare.com/datasets/detail?pageType=datasets&name=W_DPA_v1_0&id=136}{$^{\mathrm{url}}$},
\ce{C_{12}H_{26}}~\cite{zeng2020exploring}\href{https://www.aissquare.com/datasets/detail?pageType=datasets&name=C12H26_DPA_v1_0&id=140}{$^{\mathrm{url}}$} and 
\ce{HfO2}~\cite{wu2021deep}\href{https://www.aissquare.com/datasets/detail?pageType=datasets&name=HfO2_DPA_v1_0&id=145}{$^{\mathrm{url}}$}
are used for \textbf{pre-training}, while 
% HEA*\href{https://www.aissquare.com/datasets/detail?pageType=datasets&name=HEA_DPA_v1.0&id=138}{$^{\mathrm{url}}$}, 
\ce{In2Se3}~\cite{wu2021accurate}\href{https://www.aissquare.com/datasets/detail?pageType=datasets&name=In2Se3_DPA_v1.0&id=144}{$^{\mathrm{url}}$}, 
\ce{H2O}-DPLR~\cite{zhang2022deep}\href{https://www.aissquare.com/datasets/detail?pageType=datasets&name=H2O-DPLR_DPA_v1.0&id=148}{$^{\mathrm{url}}$},
\ce{H2O}-SCAN0~\cite{zhang2021modeling}\href{https://www.aissquare.com/datasets/detail?pageType=datasets&name=H2O-SCAN0_DPA_v1.0&id=128}{$^{\mathrm{url}}$},
\ce{H2O}-PBE0TS~\cite{zhang2018deep}\href{https://www.aissquare.com/datasets/detail?pageType=datasets&name=H2O_DPA_v1.0&id=149}{$^{\mathrm{url}}$} and \ce{H2O}-PBE0TS-MD~\cite{distasio2014individual}\href{https://www.aissquare.com/datasets/detail?pageType=datasets&name=H2O-PBE0TS-MD_DPA_v1_0&id=189}{$^{\mathrm{url}}$},
% SSE-PBESol~\cite{huang2021deep}\href{https://www.aissquare.com/datasets/detail?pageType=datasets&name=SSE-PBESol_DPA_v1.0&id=134}{$^{\mathrm{url}}$} and AgAu-PBED3~\cite{wang2021generalizable}\href{https://www.aissquare.com/datasets/detail?pageType=datasets&name=AgAu-PBED3_DPA_v1.0&id=154}{$^{\mathrm{url}}$} 
are utilized for \textbf{downstream} tasks. Notably, four \ce{H2O} downstream datasets (\ce{H2O}-DPLR, \ce{H2O}-SCAN0, \ce{H2O}-PBE0TS and \ce{H2O}-PBE0TS-MD) significantly differ from the pre-training dataset \ce{H2O}-PD. 
% The SSE-PBESol and AgAu-PBED3 datasets only change their functional from the pre-training datasets SSE-PBE and AgAu-PBE, respectively.
\end{itemize}

\section{Single-task benchmark of the generalizability}
\label{sm:sec:single_task}

\begin{table}
  \caption{The test accuracy of the single-task DPA-2 model compared with ANI-1x~\cite{smith2018less}. 
  The energy and force test RMSEs in kcal/mol and kcal/mol/\AA\ are respectively presented. 
  The DPA-2 model is trained with a batch size of 20 for $\sim$20.2 epochs.
  The uncertainties of the ANI-1x were estimated by an ensemble of the ANI-1x model, while we have only trained on the DPA-2 model, thus no error uncertainty is provided. 
  }
  \label{sm:tab:ani-test}
  \centering  
  \begin{tabular}{lC{2.5cm}C{2.5cm}  |  C{2.5cm}C{2.5cm}  }
    \toprule
    & \multicolumn{2}{c|}{Energy RMSE [kcal/mol]} & \multicolumn{2}{c}{Force RMSE [kcal/mol/\AA]} \\\hline
Test set        &ANI-1x         &DPA-2     &ANI-1x          &DPA-2     \\
ANI-MD          &5.94$\pm$1.48     &3.31      &4.24$\pm$0.63      &1.42      \\
DrugBank        &6.01$\pm$3.01     &4.22      &5.35$\pm$1.82      &1.56      \\
GDB07-09        &1.50$\pm$0.05     &1.06      &3.93$\pm$0.08      &1.10      \\
GDB10-13        &3.21$\pm$0.14     &1.81      &6.01$\pm$0.17      &1.75      \\
S66x8           &3.01$\pm$0.18     &1.74      &2.76$\pm$0.30      &0.98      \\
Tripeptide      &3.77$\pm$0.47     &2.62      &4.79$\pm$0.70      &1.38      \\
    \toprule
  \end{tabular}
\end{table}

The DPA-2 model is trained in a single-task manner on the ANI-1x dataset~\cite{smith2018less} with a batch size of 20 for $\sim$20.2 epochs. 
The dataset is comprised of $\sim$5.0M training and $\sim$0.5M test data. 
We have tested this DPA-2 model on the  6 test datasets provided in the Ref.~\cite{smith2018less}, i.e.~ANI-MD, DrugBank, GDB07-09, GDB10-13, S66x8, and Tripeptide. 
The RMSEs of energy and force predictions are reported in Tab.~\ref{sm:tab:ani-test} and are compared to the ANI-1x model. 
The DPA-2 model shows superior accuracy over the ANI-1x model.

\begin{table}
  \caption{The test accuracy of single-task trained models on the pre-training datasets.
    The energy and force test RMSEs are reported.
    Six models are compared, the GNO (GemNet-OC~\cite{gasteiger2022graph}), EFV2 (EquiformerV2~\cite{liao2023equiformerv2}), NequIP~\cite{batzner20223}, Allegro~\cite{musaelian2022learning}, MACE~\cite{batatia2022mace} and DPA2. 
    % \WH{plz add citation in the table caption}
    All the models are trained for 1 million steps with a batch size of 1 on a GPU with 32G memory.
    ``OOM'' indicates a CUDA out-of-memory error encountered during training, whereas ``/'' signifies an unresolved error that occurred during the training process.
    The last row shows the weighted average RMSE (WARMSE) over all the datasets. 
    The weights for the datasets are defined in Tab.~1 of the main text. 
  }
 \scriptsize
  \label{sm:tab:single_task}
  \centering  
  \begin{tabular}{L{1.68cm}R{0.56cm}R{0.56cm}R{0.56cm}R{0.56cm}R{0.56cm}R{0.56cm}  | 
 R{0.56cm}R{0.56cm}R{0.56cm}R{0.56cm}R{0.56cm}R{0.56cm}}
    \toprule
    & \multicolumn{6}{c|}{Energy RMSE [meV/atom]} & \multicolumn{6}{c}{Force RMSE [meV/\AA]} \\ \hline
    Dataset             & { GNO} & { EFV2} & {NequIP} & { Allegro} & { MACE} & { DPA2} & { GNO} & { EFV2} & {NequIP} & { Allegro} & { MACE} & { DPA2}\\
Alloy                   &       14.3    &8.5    &44.0   &21.4   &16.2  &16.8   &85.1   &62.7    &175.6    &119.4   &190.2    &125.7  \\
Cathode-P               &       1.5     &1.1    &14.3   &1.0    &2.6   &0.9    &17.9   &14.9    &14.3     &24.2    &37.8     &24.5   \\
Cluster-P               &       47.7    &34.6   &75.1   &54.8   &41.3  &31.5   &69.6   &104.4   &216.6    &174.1   &189.7    &126.0  \\
Drug                    &       40.5    &29.8   &21.6   &13.1   &/     &12.7   &93.6   &807.4   &187.2    &100.8   &/       &125.5  \\
FerroEle-P              &       1.5     &1.1    &1.1    &0.7    &2.3   &0.6    &17.9   &13.0    &23.0     &28.6    &31.7    &28.7   \\
OC2M                    &       25.0    &6.7    &97.4   &61.3   &/     &36.2   &129.1  &45.2    &226.1    &166.8   &/       &154.0  \\
SSE-PBE-P               &       2.7     &OOM    &1.6    &1.0    &1.8   & 1.4   &8.2    &OOM     &41.1     &47.8    &29.9    &50.3   \\
SemiCond                &       8.0     &3.9    &20.5   &6.8    &12.7  &5.5    &94.4   &40.8    &180.7    &146.8   &182.8   &123.6  \\
H2O-PD                  &       OOM     &OOM    &0.9    &OOM    &79.9  &0.5    &OOM    &OOM     &27.1     &OOM     &29.7    &24.7   \\
Ag$\cup$Au-PBE          &       106.0   &23.4   &42.3   &39.2   &369.1 &2.4    &8.0    &4.4     &43.8     &58.9    &34.5    &17.8   \\
Al$\cup$Mg$\cup$Cu      &       5.9     &1.9    &38.0   &18.3   &7.7   &2.1    &9.4    &5.7     &48.3     &40.6    &42.9    &19.1   \\
Cu                      &       6.1     &1.7    &6.2    &1.3    &38.8  &1.2    &5.8    &3.8     &16.7     &8.9     &13.6    &8.9    \\
Sn                      &       8.4     &5.2    &18.2   &5.6    &/     &4.1    &33.7   &19.6    &62.2     &40.2    &/       &54.4   \\
Ti                      &       44.5    &19.1   &27.6   &6.9    &8.3   &5.0    &87.9   &48.6    &137.4    &85.6    &94.2    &113.1  \\
V                       &       17.9    &5.6    &8.8    &4.2    &14.2  &4.1    &79.3   &47.4    &91.6     &82.1    &140.4   &90.8   \\
W                       &       79.1    &46.8   &20.8   &4.0    &15.6  &5.6    &81.2   &51.3    &160.4    &101.6   &181.2   &108.1  \\
C12H26                  &       135.8   &123.1  &121.4  &140.4  &81.9  &55.3   &518.7  &907.4   &715.6    &648.1   &802.3   &692.5  \\
HfO2                    &       1.2     &1.0    &1.5    &1.4    &2.3   &1.0    &16.1   &9.1     &58.8     &64.0    &14.7    &54.2   \\
\textbf{WARMSE}         &       \textbf{22.4}    &\textbf{14.0} &\textbf{36.3}   &\textbf{23.4}   &\textbf{29.1}    &\textbf{13.6}   &\textbf{74.6}   &\textbf{188.9}  &\textbf{142.5}  &\textbf{108.4}   &\textbf{112.8}  &\textbf{99.2}  \\
    \toprule
  \end{tabular}
\end{table}

We train the GemNet-OC~\cite{gasteiger2022gemnet}, Equiformer V2~\cite{liao2023equiformerv2}, NequIP~\cite{batzner20223}, Allegro~\cite{musaelian2022learning}, MACE~\cite{batatia2022mace} and DPA-2 models on the pre-training datasets using the single-task training approach,
and report the test RMSEs of energy and force in Tab.\ref{sm:tab:single_task}.
All the models are trained with a batch size of 1 for 1 million steps. 
Note that the 1 million training steps are far from enough for a large dataset like OC2M, but it is reasonable to fairly compare the performance of the models under a limited time and computational resources budget. 
The learning rate, energy/force loss prefactors, and all other training hyper-parameters follow the default values provided by corresponding implementation packages. 
More details on the training settings are found in Secs.~\ref{sm:sec:hyper-params-dpa2} and \ref{sm:sec:other-models}.
It should be noted that some packages do not handle energy bias based on element type, but rely on an overall mean energy for normalization. Training in this manner can lead to a complete lack of energy convergence in many systems. To aid these models in achieving better convergence, we subtract the element-specific energy bias obtained through least squares fitting during data preprocessing, which is automatically handled internally within the DPA-2 framework.
The RMSEs are weighted averaged to obtain the WARMSE score for an easy comparison of the overall accuracy on all the pre-training datasets. 
Here we used the same weights as those used in the multi-task pre-training, see Tab.~1 in the main text. 
From Tab.~\ref{sm:tab:single_task}, it is observed that the DPA-2 model presents the smallest energy WARMSE, and the second smallest force WARMSE among all the models. 
It is noted that among the models, DPA-2, NequIP and Allegro are conservative, but GemNet-OC and Equiformer V2 are not.

\section{Multi-task training of the DPA-2 on the pre-training dataset}
\label{sm:sec:multi_task}

\begin{table}
  \caption{The test accuracy of the single-task (ST) and multi-task (MT) DPA-2 models on the pre-training datasets. 
    The energy and force test RMSEs are reported.
    The DPA-2 MT model is trained by 8 GPUs with a total batch size of 8 for 1,000K training steps. 
    Each datum in the mini-batch is randomly sampled from the pre-training dataset with a probability proportional to the weights defined by Tab.~1 in the main text. 
    The DPA-2 ST models are trained with a batch size of 1 by the effective number of training steps calculated as weight/13.2$\times$8$\times$1,000K.
  }
  \label{sm:tab:multi_task}
  \centering  
    \begin{tabular}{l R{2.0cm}| R{1.7cm}R{1.7cm} |  R{1.7cm}R{1.7cm} }
    \toprule
                        &                &\multicolumn{2}{c|}{Energy RMSE [meV/atom]} &\multicolumn{2}{c}{Force RMSE [meV/\AA]} \\\hline
Dataset                 &Eff. tr. steps  &DPA-2 ST   &DPA-2 MT   &DPA-2 ST   &DPA-2 MT   \\
Alloy                   &1212K           &16.4       &36.5       &123.9      &169.5      \\
Cathode-P               &606K            &1.2        &3.3        &36.6       &39.8       \\
Cluster-P               &606K            &33.2       &34.4       &169.9      &162.5      \\
Drug                    &1212K           &11.9       &20.6       &117.6      &128.9      \\
FerroEle-P              &606K            &0.9        &4.4        &34.9       &44.2       \\
OC2M                    &1212K           &36.0       &29.3       &175.0      &157.6      \\
SSE-PBE-P               &606K            &1.5        &2.1        &54.8       &64.0       \\
SemiCond                &606K            &6.6        &6.5        &132.8      &131.9      \\
H2O-PD                  &606K            &0.6        &3.2        &31.0       &39.7       \\
Ag$\cup$Au-PBE          &121K            &16.7       &9.4        &32.5       &28.2       \\
Al$\cup$Mg$\cup$Cu      &182K            &11.4       &4.9        &33.9       &23.4       \\
Cu                      &61K             &3.6        &3.6        &22.7       &18.2       \\
Sn                      &61K             &12.5       &24.8       &75.6       &69.7       \\
Ti                      &61K             &26.2       &16.3       &158.5      &112.4      \\
V                       &61K             &18.9       &13.9       &133.2      &110.2      \\
W                       &61K             &20.6       &24.6       &176.1      &157.9      \\
C12H26                  &61K             &92.5       &62.5       &894.4      &710.6      \\
HfO2                    &61K             &2.4        &3.9        &107.5      &102.8      \\
\textbf{WARMSE}         & --             &\textbf{14.9}       &\textbf{18.6}       &\textbf{111.1}      &\textbf{116.3}      \\
    \toprule    
  \end{tabular}
\end{table}

We train the multi-task DPA-2 model on the pre-training dataset with a batch size of 8 for 1,000K training steps and report the test energy and force RMSEs in Tab.~\ref{sm:tab:multi_task}. The learning rate undergoes an exponential decay from an initial value of 2e-4 to a final value of 3e-8, consistent with the single-task training protocol. Additionally, the gradients across different heads on each GPU are aggregated and averaged.
% The multi-task training scheme is  effective if the validation accuracy is comparable to the single-task model on all the pre-training datasets. 
Different pre-training datasets are trained with different weights (different probability to be sampled during training), 
thus for a fair comparison, the singles-task models are trained by an effective number of training steps calculated by weight/13.2$\times$8$\times$1,000K, where 13.2 is the summation of all the dataset weights. 
All the single-task models share the same structure and the same learning rate as the multi-task model.
The test RMSEs in Tab.~\ref{sm:tab:multi_task} are weighted averaged (WARMSE) for a clear comparison over all the pre-training datasets. 
In terms of training the multi-task is more difficult than the single-task scheme, because the multi-task model should simultaneously fit all the datasets with the same model capacity and number of training steps as the single-task model. 
It is shown that the force accuracy of the multi-task DPA-2 model is almost the same as the single-task DPA-2 models, while the energy RMSE is roughly 40\% higher.
We are satisfied with this accuracy on the pre-training dataset. 
The main advantage of the multi-task model is the stronger generalizability to the datasets that are not explicitly included in the pre-training datasets. 

% In the multitask scenario, we control the total number of training steps to remain unchanged. Under this condition, we trained the DPA-2 model on 8 GPUs for 1 million steps, with the results presented in the last two columns of Table.~\ref{table:training}.
% In this highly challenging situation, there are two main difficulties: (1) For a model with a similar parameter size to a single-task model to train simultaneously on 18 datasets, it poses a significant test to the model's capability and capacity; (2) By only keeping the total training steps the same, the number of training steps allocated to each dataset is effectively reduced (for 18 datasets and 8 GPUs, if the weights are uniform, the total steps assigned to each dataset should be 1 million * 8 / 18), which poses a considerable challenge to the accuracy of individual datasets. 

% From the results, we can observe that the performance on most datasets is in line with expectations, approaching the single-task outcomes of DPA-2.
% Moreover, the accuracy of some datasets even surpasses that of the single-task results, indicating that there is mutual assistance among the datasets during the training process.

\section{Downstream learning curves}
\label{sm:sec:lcurve}

\begin{figure*}
    \includegraphics[width=1.0\textwidth]{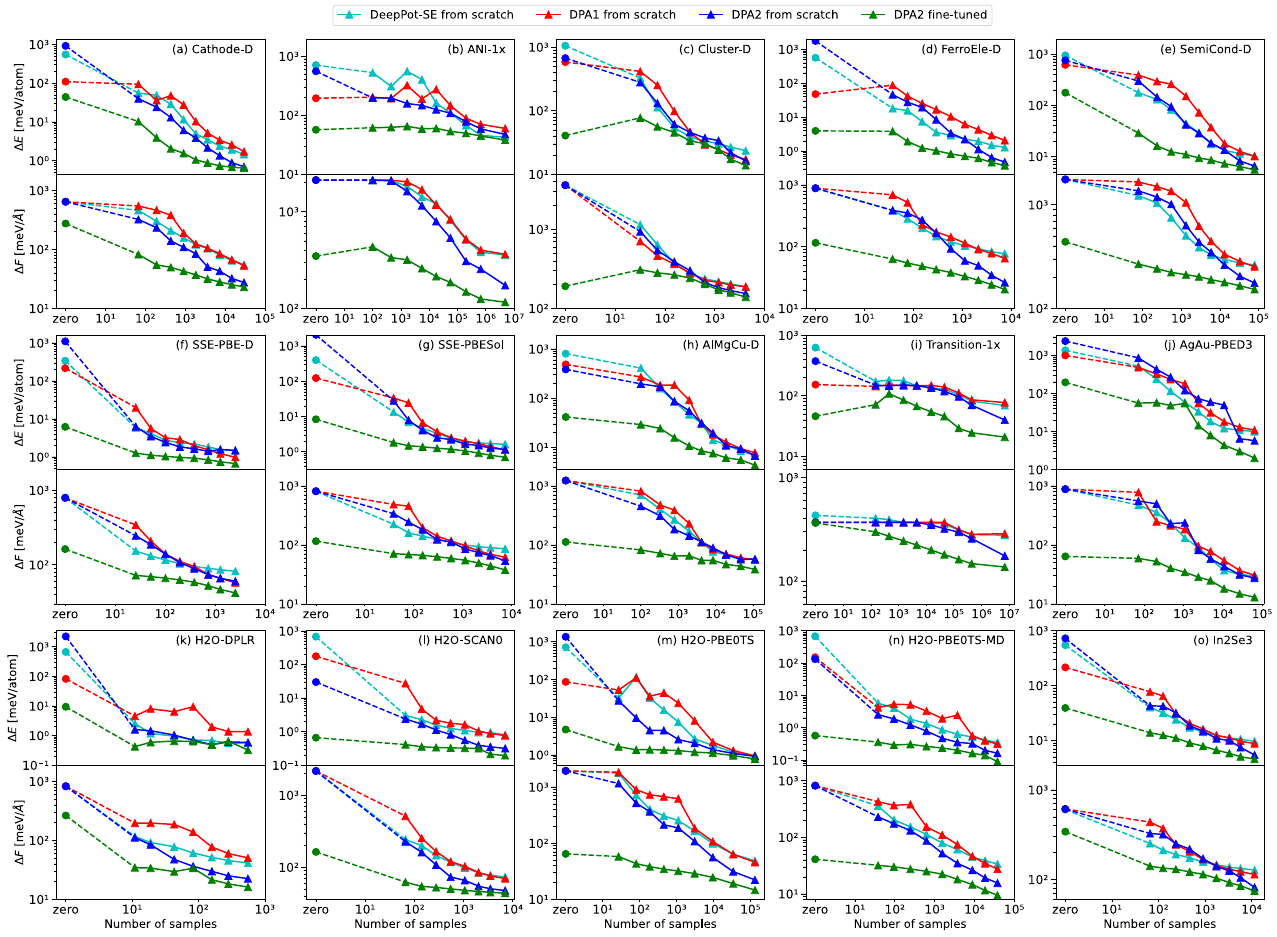}
    \centering
    \caption{Comparative analysis of sample efficiency on all downstream tasks.
    }
    \label{fig:lcurve_large}
\end{figure*}

As discussed in Sec.~2.4 of the main text, we conducted a comparative analysis of sample efficiency across all 15 downstream tasks, and illustrate the results in Fig.~\ref{fig:lcurve_large}.

\section{Choices of fine-tuning heads}

\begin{figure*}
    \includegraphics[width=0.7\textwidth]{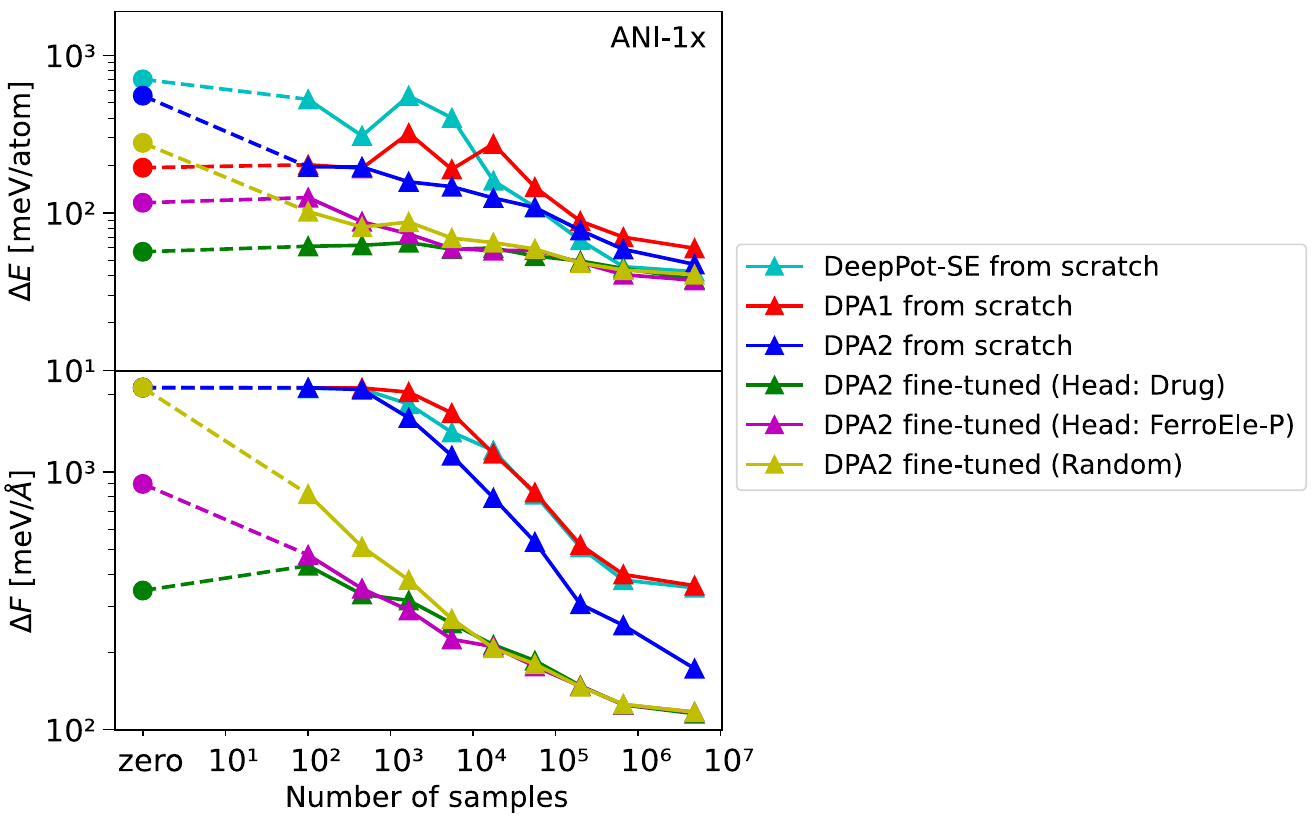}
    \centering
    \caption{
    Evaluation of various fitting head selections on the ANI-1x dataset. 
    The inclusion of a dataset name in parentheses indicates the initialization by a specific fitting head that was pre-trained on the corresponding dataset during multi-task pre-training.
    "Random" denotes the random initialization of the fitting network from scratch.
    }
    \label{fig:lcurve_head}
\end{figure*}

When fine-tuning the DPA-2 for various downstream tasks, we have the option to either select the most task-relevant head from the model's array of pre-trained heads or randomly choose a head, or even randomly initialize a new head from scratch. 
In the main text, we opted for the head that closely aligns with the characteristics of each downstream task.
To evaluate the robustness of the choice, we investigated the impact of different head selections on sample efficiency using the ANI-1x dataset, as depicted in Fig.~\ref{fig:lcurve_head}.
The results indicate that, due to the majority of parameters being pre-trained within the descriptor, the choice of fitting head had a minimal influence on the learning curve.
The learning curves of Drug and FerrorEle-P heads become identical when the sample size is larger than $\sim 10^3$. 
The learning curve of randomly initialized head converges to the other two heads after $\sim 10^4$.
Compared to the total sample size of $4.8 \times 10^6$, the influence of different fitting heads is not important.
This observation confirms the resilience of the multi-task pre-training approach.

\section{Distillation and validation by applications}

\begin{table*}
\centering
\caption{Comparison of the accuracy of few-shot fine-tuned teacher models and student models in the model distillation process across three downstream systems, along with the accuracy of DPA-1 models trained on full downstream data as a benchmark.
The units for energy and force RMSEs are expressed in meV/atom and meV/\AA, correspondingly. 
}
\label{sm:tab:table_distill}
\begin{tabular}{@{}l|ccccc|cc@{}}
\toprule
\multirow{3}{*}{Dataset} & \multicolumn{3}{c}{Teacher Model}                                                        & \multicolumn{2}{l|}{Student Model}                                              & \multicolumn{2}{l}{Full-data Model}                                           \\ \cmidrule(l){2-8} 
                         & \multicolumn{3}{c}{\begin{tabular}[c]{@{}c@{}}DPA-2 \\ few-shot fine-tuned\end{tabular}} & \multicolumn{2}{c|}{\begin{tabular}[c]{@{}c@{}}DPA-1 \\ (wo attn)\end{tabular}} & \multicolumn{2}{c}{\begin{tabular}[c]{@{}c@{}}DPA-1\\ (wo attn)\end{tabular}} \\ \cmidrule(l){2-8} 
                         & \begin{tabular}[c]{@{}c@{}}data \\ used (\%)\end{tabular}       & $\Delta$E         & $\Delta$F          & $\Delta$E                                       & $\Delta$F                                     & $\Delta$E                                     & $\Delta$F                                     \\ \midrule
H2O-PBE0TS-MD                 & 0.25                                                             & 0.3       & 30.0       & 0.4                                     & 41.7                                 & 0.4                                   & 37.7                                 \\
SSE-PBE-D                & 1.01                                                            & 1.3       & 72.0       & 3.3                                    & 101.0                                 & 1.9                                   & 96.7                                  \\
FerroEle-D                      &7.86                                                            & 1.6       & 44.7       & 2.5                                    & 99.6                                  & 1.9                                  & 105.7                                 \\ \bottomrule
\end{tabular}
\end{table*}

As outlined in Sec.~2.5 of the main text, we executed model distillation on three representative downstream tasks: H2O-PBE0TS-MD, SSE-PBE-D, and FerroEle-D.
This process was conducted in accordance with the methodologies detailed in Sec.~4.6.
% The main results are reported in Fig.~\ref{sm:fig:evaluation}.
The RMSEs of both the fine-tuned DPA-2 models (teacher models) and the distilled DPA-1 models (student models) on test data are presented in Tab.~\ref{sm:tab:table_distill}.
The DPA-1 models with the same architecture are trained on the full downstream datasets for comparison.
As is explained in the main text, in the downstream task of FerroEle-D, the pre-training dataset FerroEle-P is append to the 7.86\% subset of FerroEle-D for fine-tuning. 

% These models 

% It reveals that employing the DPA-2 model, which has been few-shot fine-tuned, as the teacher model for distilling knowledge into the DPA-1 model, endows the student DPA-1 model with the capability to match or even surpass the peak accuracy traditionally attained by the original DPA-1 model structure when trained on the full dataset.

\section{Ablation study}
\label{sm:sec:ablation}

\begin{table*}
\centering
\caption{Ablation study on the components within the $\mathrm{repformer}$ layer. This study investigates the impact of sequentially removing the respective components from left to right within the single- and pair-atom channels independently.
The table presents the average variations of the test RMSEs across all datasets during the multi-task pre-training process. 
% \WH{check if the ablation study is carried on the multi-task pre-training.}
The units for energy and force RMSEs are meV/atom and meV/Å, respectively. 
% \WH{Thus, the errors presented here are the validation error on the pre-training dataset?}
A larger change indicates a more critical contribution to the accuracy of the model.}
\begin{tabular}{@{}ccccccccc@{}}
\toprule
Structures                                                                       & \multicolumn{8}{c}{Increased RMSEs when removed sequentially $\rightarrow$}                                                                                                \\ \midrule
\multirow{3}{*}{\begin{tabular}[c]{@{}c@{}}single-atom\\ channel\end{tabular}} & \multicolumn{2}{c}{\textbf{conv}}     & \multicolumn{2}{c}{\textbf{sym\_f}} & \multicolumn{2}{c}{\textbf{sym\_g}} & \multicolumn{2}{c}{\textbf{local\_attn}} \\ \cmidrule(l){2-9} 
                                                                                 & $\Delta$E                 & $\Delta$F                 & $\Delta$E                & $\Delta$F                & $\Delta$E                & $\Delta$F                & $\Delta$E                  & $\Delta$F                   \\
                                                                                 & +5.1              & +41.2             & +0.7             & +10.4            & +17.1             & +51.3            & +14.6               & +58.4               \\ \midrule
\multirow{3}{*}{\begin{tabular}[c]{@{}c@{}}pair\\ channel\end{tabular}}          & \multicolumn{2}{c}{\textbf{prod\_f}} & \multicolumn{2}{c}{\textbf{gate}}   & \multicolumn{2}{c}{\textbf{attn}}   & \multicolumn{2}{c}{\multirow{3}{*}{}}    \\ \cmidrule(lr){2-7}
                                                                                 & $\Delta$E                 & $\Delta$F                 & $\Delta$E                & $\Delta$F                & $\Delta$E                & $\Delta$F                & \multicolumn{2}{c}{}                     \\
                                                                                 & +4.6              & +21.7              & +4.7             & +34.6            & +0.6             & +14.1             & \multicolumn{2}{c}{}                     \\ \bottomrule
\end{tabular}
\label{table:ablation}
\end{table*}

In this section, we carry out an ablation study to investigate the importance of different structures in the $\mathrm{repformer}$ layer. 
To begin with, we introduce some notations to denote specific structures. For the update of the single-atom representation $f_i^{2,l}$, we use $\mathrm{\textbf{conv}}$ to signify $\frac{1}{N_{r_c}^m}\sum_{j\in N_{r_c}(i)} w_{ij} g_{ij}^{2,l} f_j^{2,l}$, which is the convolution part in Eq.~(25) of the main text. 
Furthermore, $\mathrm{\textbf{sym\_f}}$ and $\mathrm{\textbf{sym\_g}}$ represent the substructures $\mathrm{symm}\big( f_j^{2,l}, h_{ij}^{2,l}\big)$ and $\mathrm{symm}\big( g_{ij}^{2,l}, h_{ij}^{2,l}\big)$ in Eq.~(25), respectively. 
Meanwhile, $\mathrm{\textbf{local\_attn}}$ denotes the local attention operator in Eq.~(24).
With regard to the update of pair-atom representation $\hh^{2,l}_{ij}$, $\mathrm{\textbf{prod\_f}}$ stands for the element-wise product $w_{ij} \textrm{linear} (f_i^{2,l} \odot f_j^{2,l}) $ in Eq.~(29), and we divide the $\mathrm{gated\_attn} \big(g_{ij}^{2,l}, h_{ij}\big)$ in Eq.~(29) into $\mathrm{\textbf{gate}}$ and $\mathrm{\textbf{attn}}$ to represent the $ \sum_{\delta} \rr_{ij,\delta} \rr_{ik,\delta}$ in Eq.~(32) and other parts in Eq.~(30)--(32), respectively.

We conducted separate ablation studies on the single-atom representation $f_i^{2,l}$ and pair-atom representation $\hh^{2,l}_{ij}$ within the $\mathrm{repformer}$ layer. 
Tab.~\ref{table:ablation} presents the results, where we sequentially remove the corresponding structures from left to right and observe the changes in the average RMSE across all datasets during the multi-task pre-training process. 
It is evident that the removal of certain structures leads to varying degrees of increase in RMSE. For the single-atom representation, all structures except $\mathrm{\textbf{sym\_f}}$ exhibit substantial importance. 
Conversely, for the pair-atom representation, the $\mathrm{\textbf{gate}}$ component appears to be more significant than the attention operator $\mathrm{\textbf{attn}}$ itself.

\section{Energy conservation}
\label{sm:sec:conserv}

\begin{figure*}
    \includegraphics[width=1.0\textwidth]{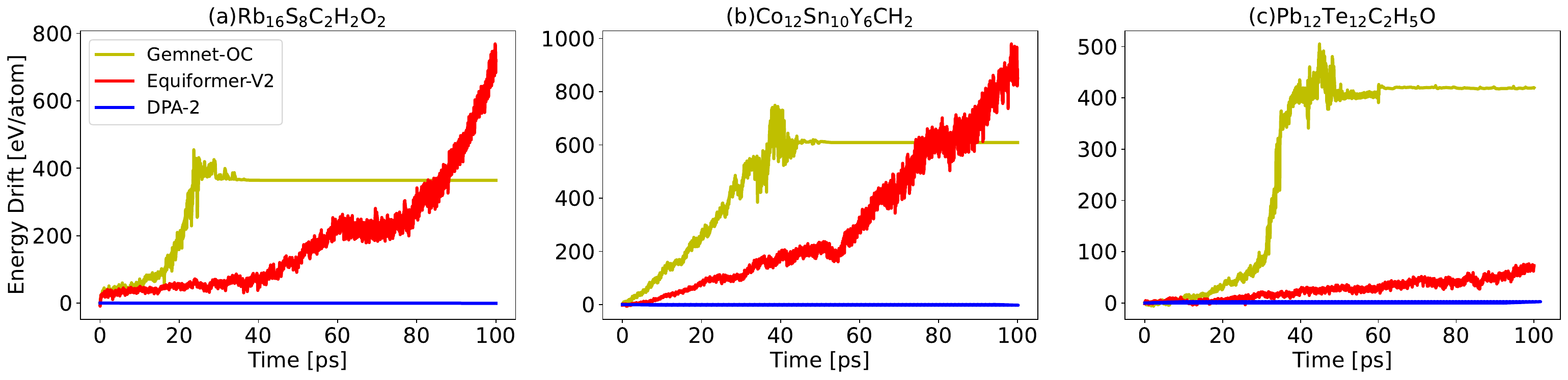}
    \centering
    \caption{Drift of total energy in 100~ps-long NVE MD simulations for Gemnet-OC, Equiformer-V2, and DPA-2 models, where (a-c) are three randomly selected structures from the OC2M dataset, and the simulation temperature is initialized to 330 K.
    }
    \label{fig:NVE_MD}
\end{figure*}

To illustrate the energy conservation of the DPA-2 model and the energy drift caused by non-conservative models, we conducted NVE MD simulations on three randomly chosen OC2M structures, as depicted in Fig.~\ref{fig:NVE_MD}. 
We downloaded the GemNet-OC and Equiformer-V2 model checkpoints from their official websites, with reported energy MAEs of 0.286 eV and 0.236 eV, and force MAEs of 0.026 eV/\AA and 0.016 eV/\AA, respectively.
The training parameters for GemNet-OC included a batch size of 16 over 10 million steps, while Equiformer-V2 was trained with a batch size of 4 for 15 million steps.
The DPA-2 model is derived from the OC2M single-task training, with the same settings in Sec.~\ref{sm:sec:single_task}. 
It was trained using a batch size of 1 for 1 million steps, which is distant from fair convergence, as indicated by an energy MAE of 0.912 eV and a force MAE of 0.116 eV/\AA. 
This model serves merely as an illustrative case.
% \WH{These numbers are not comparable to the table S2. either provide DPA errors in the same sense or test the GNO and EFV2 models in our sense!}
% The Gemnet-OC and Equiformer-V2  models are more accurate than the DPA models as they were trained much longer. 
Despite the high energy and force accuracy of the non-conservative models (Gemnet-OC and Equiformer-V2), their lack of energy conservation results in a significant drift in the total energy during the simulation. 
In contrast, the DPA-2 simulations do not exhibit any energy drift.

\section{Hyper-parameters of the DPA-2 model}
\label{sm:sec:hyper-params-dpa2}

In our study, we employed consistent descriptor configurations for both single-task and multi-task training using the DPA-2 framework. Specifically, we configured the $\mathrm{repinit}$ layer with a cutoff radius ($r_c^0$) of 9.0 \AA~and a maximum neighbor count in $N_{r_c^0}^m$ of 120. 
The dimension of the input single-atom representation is 8. 
The MLP used for embedding $\mathrm{concat}(f_i^0, f_j^0, g_{ij}^0)$ (Eq.(16) in the main text) consists of three layers, each of which has 25, 50, and 100 neurons respectively.
The dimension of the pair representations $g^{rt}_{ij}$ is 100. 
The first 12 dimensions is preserved in the split operation (Eq.~(20)) in the symmetrization operator, so the output dimension of single-atom representation is 1200. 
% \WH{ZD plz check if previous sentence is correct. }
For the $\mathrm{repformer}$ layers, we chose to use 12 layers, with a cutoff radius ($r_c^1$) of 4.0~\AA\ and a $N_{r_c^1}^m$ of 40.
We designated the dimensions of the single-atom and pair-atom feature representations to be 128 and 32, respectively. 
Furthermore, within each local multi-head self-attention layer, we set the hidden dimension to 128. 
% \WH{how many heads?}
For the gated multi-head self-attention layers, we opted for a dimension of 32 and configured both types of layers with four attention heads. The fitting network is comprised of a three-layer MLP with each layer containing 240 neurons. 

Regarding the distillation model, we are at liberty to employ parameters specific to downstream tasks. Take the distillation process on the H2O-PBE0TS-MD dataset as an example, we implemented the DPA-1 model architecture excluding the attention layers. We established a cutoff radius ($r_c^0$) of 6.0 Å and capped the maximum number of neighbors at 120. Additionally, the MLP configuration for the embedding and fitting networks mirrors the settings used in the $\mathrm{repinit}$ layer of the DPA-2 framework.

For the single-task model, the total parameter count is approximately 5.15 million, with the descriptor accounting for nearly 5 million parameters and the fitting network utilizing around 0.15 million. In comparison, the multi-task model, trained on 18 pre-training datasets, boasts a total parameter count of about 7.68 million. This total includes 5 million descriptor parameters, augmented by an extra 2.68 million parameters distributed across the 18 fitting networks. The distilled model exhibits a total parameter count of around 0.53 million.
% \WH{Mention the total number of parameter}

% \WH{Mention the hyper parameters of the DPA-1 model used in distillation.}

As for our training strategy, we initiated the learning rate at 2e-4, applying an exponential decay after every 5,000 steps, and ultimately reducing it to 3e-8 upon reaching 1 million training steps. The prefactors for both energy and force were adjusted in conjunction with the learning rate. Specifically, the energy prefactor was scaled from 0.02 to 1, while the force prefactor underwent a change from 1,000 to 1.

\section{Training details of GemNet-OC/EquiformerV2/Allegro/NequIP/MACE}
\label{sm:sec:other-models}

% I restrict the length of hash commit to 7, otherwise the latex cannot perfectly format the very long strings. 
For GemNet-OC and EquiformerV2, the training code is from the main branch of the GitHub repository \url{https://github.com/Open-Catalyst-Project/ocp} 
(commit hash: \texttt{9bc9373}), 
% (commit hash: 9bc93735aa3504952e49a46db94d4c7efafc0129), 
and the training parameters are from the file s2ef/all/gemnet/gemnet-dT.yml and configs/s2ef/2M/equiformer\_v2/equiformer\_v2\_N@12\_L@6\_M@2.yml in the repository except for the batch size set to 1 for a fair comparison. 
For Allegro, the training code is from the main branch of the GitHub repository \url{https://github.com/mir-group/allegro} 
(commit hash: \texttt{22f673c}), 
% (commit hash: 22f673c565d148ae8e9394443baa9f5b5716c8e7), 
and the training parameters are from the file configs/example.yaml in the repository. The batch size is also set to 1, and the learning rate is set to 0.005 which is tested to be optimal when 1 is used for batch size. 
For NequIP, the training code is from the main branch of the GitHub repository \url{https://github.com/mir-group/nequip} 
(commit hash: \texttt{dceaf49}, tag: v0.5.6), 
% (commit hash: dceaf49100987d1dff87873c28a5b1551cb3b76c, tag: v0.5.6), 
and the training parameters are from the file configs/example.yaml in the repository. The batch size is also set to 1, and the learning rate is set to 0.002 which is tested to be optimal when 1 is used for batch size.
% \WH{provide the access date for GemNet-OC and EquiformerV2, and the commit hashes of allegro and nequip.}
For MACE, the training code is from the main branch of the GitHub repository \url{https://github.com/ACEsuit/mace} (commit hash: \texttt{b76a2a9}), 
% (commit hash: b76a2a947576d8b10fd5f72ccf2053aebaed96f7), 
and the training parameters are from the medium settings of MACE-MP-0 (\url{https://github.com/ACEsuit/mace-mp/blob/main/mace_mp_0/2023-12-03-mace-128-L1.sh}. For a fair comparison, the batch size is set to 1, and the maximum number of epochs is adjusted to encompass just over 1 million batches for each dataset.

For the datasets we use, the contributions to the total energy of different types of atoms may differ significantly, which leads to a large variance in energy. Training on the dataset directly with GemNet-OC, EquiformerV2, Allegro or NequIP will cause the energy loss to remain high. We preprocess the dataset by subtracting the energy bias for each atom from the total energy, where the energy bias is determined by a least-square fitting of the energies in the training data. Although MACE automatically performs this adjustment internally, this preprocessing step ensures no negative impact on the results. 
For GemNet-OC, systems with fewer than 3 atoms are excluded otherwise errors will occur during the training process. 
For Allegro, systems with fewer than 2 atoms are excluded otherwise Allegro will complain about the lack of neighbors. 
For NequIP, as 4.0 is used as the cutoff radius, systems with a distance between the two closest atoms larger than 4.0 are excluded to avoid errors. 
None of GemNet-OC, EquiformerV2, Allegro, or NequIP natively support non-periodic boundary condition systems as training data. For such systems, we apply a cubic box with a side length greater than the distance between the two furthest atoms in the system plus the cutoff radius.
For MACE, the training data utilized is identical to that of DPA-2. However, we faced significant challenges when training MACE on larger datasets. To address this, we employed the conversion process in the README of GitHub repository \url{https://github.com/ACEsuit/mace} to transform the problematic datasets into optimized and recommended HDF5 files. Despite this effort, some training sessions still failed. Specifically, for the OC2M dataset, the training process stalled at the very beginning, while for the Sn dataset, it consistently crashed after several epochs. Additionally, for the Drug dataset, the conversion process became unresponsive and produced no error logs, even on a machine with over 200GB of memory.
The maximum memory required by the Allegro and NequIP programs during the data processing and training process also increases as the dataset size increases. 
For the OC2M dataset, Allegro and NequIP are unable to complete the data processing on a machine with 155GB of memory. 
We divide the OC2M dataset randomly into 4 subsets and train Allegro or NequIP on them sequentially.

\end{document}